\documentclass[10pt,a4paper]{article}
	
\usepackage{a4wide}
\usepackage{amsmath}
\usepackage{amssymb}
\usepackage{amsfonts}
\usepackage{epsfig}
\usepackage{color}
\usepackage{subfigure}
\usepackage{exscale}
\usepackage{float}
\usepackage{bbm}
\usepackage[numbers,sort&compress]{natbib}

\newcommand{\Z}{{\mathbb{Z}}}
\newcommand{\R}{{\mathbb{R}}}

\makeatletter
\@addtoreset{equation}{section}
\makeatother

\title{Ultracold Quantum Gases and Lattice Systems: \\
Quantum Simulation of Lattice Gauge Theories
\footnote{Invited contribution to the ``Annalen der Physik'' topical issue 
``Quantum Simulation'', \newline 
guest editors: R.\ Blatt, I.\ Bloch, J.\ I.\ Cirac, and P.\ Zoller.}}

\author{U.-J.\ Wiese$^{1,2}$ 
\\ \\
$^1$ Albert Einstein Center for Fundamental Physics \\
Institute for Theoretical Physics, Bern University \\
Sidlerstrasse 5, CH-3012 Bern, Switzerland \\ \\
$^2$ Center for Theoretical Physics, Massachusetts Institute of Technology \\
77 Massachusetts Avenue, Cambridge, Massachusetts, U.S.A. \\ \\}

\begin{document} 

\maketitle

\begin{abstract} \normalsize

Abelian and non-Abelian gauge theories are of central importance in many areas
of physics. In condensed matter physics, Abelian $U(1)$ lattice gauge theories 
arise in the description of certain quantum spin liquids. In quantum information
theory, Kitaev's toric code is a $\Z(2)$ lattice gauge theory. In particle 
physics, Quantum Chromodynamics (QCD), the non-Abelian $SU(3)$ gauge theory of 
the strong interactions between quarks and gluons, is non-perturbatively 
regularized on a lattice. Quantum link models extend the concept of lattice 
gauge theories beyond the Wilson formulation, and are well suited for both
digital and analog quantum simulation using ultracold atomic gases in optical 
lattices. Since quantum simulators do not suffer from the notorious sign 
problem, they open the door to studies of the real-time evolution of strongly 
coupled quantum systems, which are impossible with classical simulation methods.
A plethora of interesting lattice gauge theories suggests itself for quantum 
simulation, which should allow us to address very challenging problems, ranging 
from confinement and deconfinement, or chiral symmetry breaking and its 
restoration at finite baryon density, to color superconductivity and the 
real-time evolution of heavy-ion collisions, first in simpler model gauge 
theories and ultimately in QCD.

\end{abstract}

\newpage

\section{Introduction}

Gauge theories play an important role in many areas of physics. In particle 
physics, Abelian and non-Abelian gauge fields mediate the fundamental strong
and electroweak forces between quarks, electrons, and neutrinos. In atomic and
molecular physics, electromagnetic Abelian gauge fields are responsible for the 
Coulomb forces that bind electrons to atomic nuclei. In condensed matter 
physics, besides the fundamental electromagnetic field, effective gauge fields 
may emerge dynamically at low energies \cite{Lev05}. For example, in quantum 
Hall systems, statistical gauge fields endow Laughlin quasi-particles with anyon
statistics. Some quantum spin liquids \cite{Her04}, which may arise in 
geometrically frustrated antiferromagnets, can be described by quantum dimer 
models \cite{Rok88,Moe02} which are $U(1)$ lattice gauge theories. Other quantum
spin liquids have Abelian $\Z(2)$ or non-Abelian $SU(2)$ symmetry, and Kitaev's 
toric code \cite{Kit03} provides an example of a $\Z(2)$ lattice gauge theory 
relevant in quantum information theory. Furthermore, universal topological 
quantum computation is based on non-Abelian Chern-Simons gauge theories
\cite{Nay08}.

Gauge theories reflect a redundancy in or description of Nature. 
When we use vector potentials to talk about magnetic fields, we introduce
unphysical degrees of freedom, which ultimately decouple thanks to a local 
gauge symmetry. Similarly, when a condensed matter theorist factorizes an 
electron field into a charged spinless boson and a neutral fermion carrying the
spin, a phase ambiguity arises that again manifests itself as a gauge symmetry. 
While one may speculate whether Nature herself uses redundant variables at the 
ultimate cut-off scale, it is not surprising that gauge fields dominate the
low-energy domain. In particular, Abelian gauge fields can exist in a Coulomb 
phase, which provides us with naturally massless photons that mediate the 
electromagnetic interaction between charges at arbitrarily large distances. 
Non-Abelian gauge theories \cite{Yan54,tHo72}, on the other hand, are confining 
and give rise to massive particles. For example, in Quantum Chromodynamics (QCD)
\cite{Gro73,Pol73,Fri73} with the non-Abelian gauge group $SU(3)$, massless 
gluons and almost massless quarks are turned into massive hadrons, including 
protons and neutrons. However, thanks to the property of asymptotic freedom, 
the hadron masses are exponentially small compared to the ultimate cut-off 
scale (e.g.\ the Planck scale), and protons and neutrons hence 
also naturally participate in the low-energy physics. The robustness of gauge 
invariance at low energies, even if it is violated at the cut-off scale, has
been recognized a long time ago \cite{Foe80}. When gauge theories
undergo the Higgs mechanism, the gauge bosons pick up a mass at the scale of the
vacuum expectation value of the Higgs field. Particle physicists are puzzled by
the fact that the electroweak gauge bosons W and Z as well as the Higgs boson
exist far below the Planck scale, because the large mass hierarchy seems to 
require unnatural fine-tuning. In condensed matter physics, man-made 
fine-tuning happens on a daily basis. For example, by tuning a sophisticated 
material to a deconfined quantum critical point, one may hope to 
liberate an effective gauge boson that would otherwise remain confined and thus 
unobservable \cite{Sen04}.

The omnipresence of gauge fields confronts us with very rich low-energy 
dynamics that are often difficult to understand. In particular, in strongly
coupled materials, such as frustrated quantum magnets, high-temperature
superconductors, or dense nuclear matter, perturbative analytic methods fail,
and one must resort to numerical calculations. Despite tremendous successes of 
Monte Carlo simulations in condensed matter and particle physics, these problems
remain largely intractable, due to very severe sign problems which prevent the 
importance sampling method underlying classical and quantum Monte Carlo. Since
the dimension of the Hilbert space grows exponentially with the size of a 
quantum system, simulating it on a classical computer is in general very 
difficult. While some sign problems are NP-complete \cite{Tro05}, and thus 
believed to be practically unsolvable, the above mentioned sign problems arising
in condensed matter and particle physics are probably not of this sort. Still, 
it may require more than a classical computer to deal with them.

Since the ground-breaking experimental realization of Bose-Einstein 
condensation \cite{And95,Dav95}, the field of atomic, molecular, and optical 
physics has undergone an impressive rapid development. In particular, the 
degree to which ultracold atomic systems can be engineered and controlled is 
truly remarkable. Shor was first to show that, if it existed, a quantum 
computer would vastly outperform classical computers in the task of prime 
factorization \cite{Sho97}. Very early on, Cirac and Zoller realized 
theoretically that trapped ions could be used for quantum computation 
\cite{Cir95}. The prospect of perhaps being able to build a quantum computer 
spurred the development of algorithms for the envisioned machines. 
It was soon realized that cold atoms in optical lattices can also 
be used as quantum simulators. In this way, Feynman's vision \cite{Fey82} of 
simulating complicated physical systems by other well-controlled quantum 
systems is becoming a reality. Quantum simulators \cite{Cir12} are special 
purpose quantum computers which are used as digital \cite{Llo96} or analog 
\cite{Jak98} devices to simulate, for example, strongly coupled quantum systems 
relevant in condensed matter physics. Quantum simulators have been constructed 
using ultracold atoms in optical lattices \cite{Lew12,Blo12}, trapped ions 
\cite{Bla12}, photons \cite{Asp12}, or superconducting circuits on a chip 
\cite{Hou12}. The D-wave One Rainer chip is a $4 \times 4$ aray of unit-cells 
each hosting 8 superconducting flux qubits with programmable couplings 
connecting numerous pairs of qubits. Recently, a 108 qubit system has been used 
to run a finite-temperature variant of the quantum adiabatic algorithm 
\cite{Far00} on random instances of an Ising spin glass Hamiltonian 
\cite{Boi13}. A comparison with a simulated classical and quantum annealer led 
the authors of this study to conclude that the D-wave device indeed performs 
quantum rather than classical annealing, however, as yet without achieving any 
speed-up compared to classical computation for the problem under consideration. 
Since it utilizes the superposition of quantum phases in its hardware, a quantum
simulator does not suffer from the notorious sign problem. From a condensed 
matter perspective, this is extremely interesting because highly non-trivial 
many-body systems, such as geometrically frustrated quantum antiferromagnets
\cite{Dua03}, various spin liquids, or high-temperature superconductors 
\cite{Lee06} can perhaps be quantum simulated, despite the 
fact that the sign problem prevents numerical simulations on classical 
computers. It was a significant breakthrough, when a non-trivial quantum phase 
transition separating a Mott insulator (with localized particles) from a 
superfluid, was first quantum simulated by implementing the Bose-Hubbard model 
with cold atoms in an optical lattice \cite{Gre02}. The parameters of the 
system are controlled by varying the depth of the optical potential generated 
by appropriately tuned laser beams. Since the Bose-Hubbard model does not suffer
from a sign problem, it can also be simulated on a classical computer. Using 
quantum Monte Carlo, in this way the quantum simulator has been accurately 
validated in comparison to the cold atoms experiments \cite{Tro10}. Thus it has 
been demonstrated explicitly that accurate quantitative experimental control of 
the theoretical Bose-Hubbard model has indeed been achieved.

It is natural to ask whether one can also quantum simulate relativistic field 
theories. A field theory quantum simulator should allow us to control a 
large number of strongly coupled field degrees of freedom, by engineering 
appropriate local Hamiltonians. It should enable flexible initial state 
preparation and precise subsequent time-evolution, followed by final state 
detection. As usual for quantum systems, by repeating identically prepared 
experiments many times, one obtains physical results by averaging over them. 
Quantum simulator constructions already exist for several bosonic 
\cite{Ret05,Hor10,Jor12} and fermionic \cite{Cir10,Ber10,Boa11,Maz12} 
field theories. Quantum simulators have also been constructed for quantum
particles interacting with classical gauge fields. A most exciting development 
are synthetic background gauge fields, in order to access, for example, the 
fractional quantum Hall effect or topological insulators with atoms. Such 
synthetic gauge fields can mimic an external magnetic field 
\cite{Jak03,Ost05,Rus05,Gol09,Lin09,Lin11,Coo11,Ade11,Dal11,Edm13}. 

In particle physics, one is dealing, e.g., with $SU(3)$ gluon gauge fields, 
which are not classical background fields but dynamical quantum fields. When
coupled to quark fields, their non-perturbative quantum physics is very 
successfully described by Wilson's lattice QCD 
\cite{Wil74,Kog79,Kog83,Cre85,Mon94,DeG06,Gat10}. Still, some 
problems remain intractable with this method, because again severe sign 
problems arise. For example, the hot quark-gluon plasma that is generated in 
heavy ion collisions at the Relativistic Heavy Ion Collider (RHIC) in Brookhaven
or at the Large Hadron Collider (LHC) at CERN undergoes an intriguing real-time 
evolution that is practically impossible to derive from QCD first principles. 
Even staying within equilibrium thermodynamics, the critical endpoint of the 
chiral transition line in the QCD phase diagram, which will be investigated 
experimentally at the Facility for Antiproton and Ion Research (FAIR) at GSI 
Darmstadt, is difficult to determine accurately using lattice QCD
\cite{Ris04,Fuk11} . Similarly, due to a severe sign problem, the deep 
interior of neutron stars, which may contain color superconducting quark matter 
\cite{Raj00,Alf08} can currently not be addressed non-perturbatively from QCD 
first principles either. While it is difficult to predict when full QCD quantum 
simulations with cold atoms might become experimentally feasible, it is timely 
to work towards this long-term goal \cite{Bue05,Kap11,Zoh11,Szi11,Liu12,Zoh12,
Tag12,Ban12,Ban13,Zoh13,Tag13,Zoh13a,Zoh13b}.

What are realistic short-term goals? Obviously, a quantum simulator should 
aim at those problems that are inaccessible to classical simulation methods
(but still verifiable for certain small-scale instances). This includes all 
problems involving real-time evolution or the physics at high baryon density. 
Unlike classical computers, initially quantum simulators will not be precision 
instruments. In particular, they lack automatic error correction. One should 
hence concentrate on robust, fault-tolerant phenomena of a qualitative nature, 
such as the presence or absence of specific phases of strongly coupled 
high-density matter in Abelian and non-Abelian gauge theories, or specific 
events that may happen or not happen during real-time evolution. Since one is
interested in phenomena inaccessible to analytic perturbative calculations, one
will deal with lattice gauge theories, for which continuous space is replaced by
a discrete grid of lattice points. Rather than aiming directly at the continuum
limit of vanishing lattice spacing, it is more realistic and still extremely
interesting to first address the strong coupling lattice dynamics.

In this review, we will approach gauge theories from a particle physics point of
view, but we will also draw connections to condensed matter physics and 
quantum information theory. In section 2, we begin with the simpler case of 
Abelian gauge theories. In order to address non-perturbative effects, we
regularize the theory on a lattice, first following the classical Wilson 
formulation \cite{Wil74}, and then extending the concept of lattice gauge 
theories to quantum link models \cite{Hor81,Orl90,Cha97,Bro99}. The resulting 
Abelian gauge theories resemble models of condensed matter physics and quantum 
information theory. In section 3, we present various constructions of quantum 
simulators for Abelian gauge theories, and discuss possible experimental 
realizations. In section 4, we then proceed to non-Abelian gauge theories, first
introducing them in the framework of Wilson's lattice gauge theory, and again
extending them to quantum link models. This gives rise to the alternative 
D-theory formulation of quantum field theory \cite{Bro04}, which is well suited 
for quantum simulation with ultracold atoms. In section 5, we present several 
quantum simulator constructions of interesting non-Abelian lattice gauge 
theories. Finally, in section 6, we summarize the current status of gauge theory
quantum simulators, give an outlook on possible developments in the foreseeable
future, and conclude by discussing what will need to be done to ultimately 
facilitate quantum simulations of QCD.

\section{Abelian Gauge Theories}

In this section we consider Abelian gauge theories, starting from classical
electrodynamics, and then proceeding from the quantum mechanics of a single 
particle in a classical electromagnetic background field to the quantum field
theory of dynamical Abelian gauge fields. Quantum Electrodynamics (QED) is 
formulated non-perturbatively, first using Wilson's lattice regularization. The 
concept of Abelian lattice gauge fields is then extended to quantum link models,
which are well-suited for quantum simulation using ultracold matter. Both 
digital and analog quantum simulator constructions for $\Z(2)$ and $U(1)$ 
lattice gauge theories, with and without matter fields, are presented and 
related experiments are discussed.
 
\subsection{Classical Abelian Gauge Fields}

The most familiar gauge theory is classical electrodynamics, the theory of 
electric and magnetic fields $\vec E(\vec x,t)$ and $\vec B(\vec x,t)$, which 
are functions of space and time that obey Maxwell's equations
\begin{eqnarray}
&&\vec \nabla \cdot \vec E(\vec x,t) = \rho(\vec x,t), \ 
\vec \nabla \times \vec E(\vec x,t) + 
\partial_t \vec B(\vec x,t) = 0, \nonumber \\
&&\vec \nabla \cdot \vec B(\vec x,t) = 0, \
\vec \nabla \times \vec B(\vec x,t) - 
\partial_t \vec E(\vec x,t) = \vec j(\vec x,t),
\end{eqnarray}
where $\rho(\vec x,t)$ and $\vec j(\vec x,t)$ are the electric charge and 
current densities. Throughout this review, we are using natural units in which 
$c = \hbar = 1$. In order to ensure the absence of magnetic monopoles, i.e.\ 
$\vec \nabla \cdot \vec B(\vec x,t) = 0$, it is convenient to introduce a
vector potential $\vec A(\vec x,t)$.  Similarly, in order to guarantee that the
other homogeneous Maxwell equation is automatically satisfied as well, one also
introduces a scalar potential $\Phi(\vec x,t)$ and one writes
\begin{equation}
\vec E(\vec x,t) = 
- \vec \nabla \Phi(\vec x,t) - \partial_t \vec A(\vec x,t), \
\vec B(\vec x,t) = \vec \nabla \times \vec A(\vec x,t).
\end{equation}
The representation of electromagnetic fields in terms of scalar and vector
potentials has a high degree of redundancy. In particular, the physical fields
$\vec E(\vec x,t)$ and $\vec B(\vec x,t)$ remain invariant under arbitrary gauge
transformations $\alpha(\vec x,t)$,
\begin{equation}
\Phi(\vec x,t)' = \Phi(\vec x,t) + \partial_t \alpha(\vec x,t), \
\vec A(\vec x,t)' = \vec A(\vec x,t) - \vec \nabla \alpha(\vec x,t).
\end{equation}

\subsection{Charged Quantum Particle in a Classical Electromagnetic Background 
Field}

While in classical physics the introduction of scalar and vector potentials is 
mostly a mathematical convenience, it becomes unavoidable when we want to write
down the Schr\"odinger equation,
\begin{equation}
i D_t \Psi(\vec x,t) = - \frac{1}{2 m} \vec D \cdot \vec D \Psi(\vec x,t),
\end{equation}
for a quantum mechanical point particle of mass $m$ and charge $-e$, 
propagating in a classical electromagnetic background field. Here 
\begin{equation}
D_t \Psi(\vec x,t) = \partial_t \Psi(\vec x,t) 
- i e \Phi(\vec x,t) \Psi(\vec x,t), \
\vec D \Psi(\vec x,t) = \vec \nabla \Psi(\vec x,t) + 
i e \vec A(\vec x,t) \Psi(\vec x,t),
\end{equation}
are covariant derivatives. When the particle's wave function is gauge 
transformed to
\begin{equation}
\label{wavef}
\Psi(\vec x,t)' = \exp\left[i e \alpha(\vec x,t)\right] \Psi(\vec x,t),
\end{equation}
the covariant derivatives transform as
\begin{equation}
D_t \Psi(\vec x,t)' = 
\exp\left[i e \alpha(\vec x,t)\right] D_t \Psi(\vec x,t), \quad
\vec D \Psi(\vec x,t)' = 
\exp\left[i e \alpha(\vec x,t)\right] \vec D \Psi(\vec x,t),
\end{equation}
such that the Schr\"odinger equation transforms gauge covariantly.

\subsection{Lattice Fermions Hopping in a Classical Electromagnetic 
Background Field}

In relativistic particle physics, not only the electromagnetic field but also 
charged matter, for example, electrons and positrons are described by a field. 
This is in contrast to a single non-relativistic particle, which is just 
described by its quantum mechanical wave function. Relativistic electrons and 
positrons indeed are not point ``particles'', but arise as quantized excitations
(sometimes called ``wavicles'') of the relativistic Dirac field, which is
coupled to the electromagnetic photon field. Like other interacting field 
theories, Quantum Electrodynamics (QED) --- the field theory of electrons, 
positrons, and photons --- suffers from ultraviolet divergences, which are 
removed in the process of regularization and subsequent renormalization. In 
perturbation theory, one regularizes individual Feynman diagrams. In order to 
define gauge theories beyond perturbation theory, Wilson has regularized them 
on a lattice that replaces continuous space (and usually even space-time) by a 
regular cubic grid. The lattice spacing $a$ serves as an ultraviolet momentum
cut-off $\frac{1}{a}$. In the process of renormalization, the coupling constants
become functions of $a$ and are properly tuned to reach the continuum limit 
$a \rightarrow 0$. In order to familiarize ourselves with the lattice 
regularization in a simple setting, we first consider staggered lattice 
fermions hopping in a classical electromagnetic background field.

Unlike in typical lattice calculations, where one works with a path integral
formulation on a $(d+1)$-dimensional Euclidean space-time lattice, here we work 
in a Hamiltonian formulation in which time remains real and continuous, while 
the $d$-dimensional space $\R^d$ is replaced by a cubic lattice that
consists of points $x = (x_1,x_2,\dots,x_d)$, $x_k = n_k a$, $n_k \in \Z$. 
Putting relativistic fermions on a lattice is a subtle issue. In particular, 
when the Dirac equation is naively discretized, one faces the 
fermion doubling problem. Instead of one physical fermion, one encounters 
additional $2^d-1$ species \cite{Nie81}. In his original formulation of lattice 
QCD, Wilson has explicitly removed the unwanted doubler fermions by giving them 
a large mass at the order of the cut-off $\frac{1}{a}$. Through 
renormalization the physical fermion then also receives a large mass, which 
must be fine-tuned unnaturally, in order to obtain a light physical Dirac 
fermion with a mass far below the lattice cut-off. In the mean time, the fermion
doubling problem has been solved elegantly by Kaplan's domain wall fermions
\cite{Kap92}, which reside in an additional spatial dimension. Here we will use 
a simple way of bypassing the fermion doubling problem, first suggested by 
Susskind \cite{Sus77}. Via 
a process known as spin diagonalization, the additional fermionic degrees of 
freedom are then reduced and partly reinterpreted as fermion ``flavors''. Here 
we choose staggered fermions because they provide the simplest realization 
of relativistic massless lattice fermions.

Staggered fermions are described by anti-commuting creation and annihilation 
operators, $\psi_x^\dagger$, $\psi_x$, associated with the lattice sites $x$,
$\{\psi_x^\dagger,\psi_y\} = \delta_{xy}$,
$\{\psi_x^\dagger,\psi_y^\dagger\} = \{\psi_x,\psi_y\} = 0$.
Unlike standard Dirac fermions, staggered fermions have no additional degrees 
of freedom. In particular, the spin is encoded in the spatial 
position $x$. From a condensed matter physics perspective, staggered fermions 
thus look like spinless fermions. Free staggered fermions of mass $m$, hopping 
between neighboring lattice sites $x$ and  $y = x + \hat k$ (where $\hat k$ is 
a vector of length $a$ in the spatial 
$k$-direction) are described by the Hamiltonian
\begin{equation}
\label{Hfree}
H_{\text{free}} = - t \sum_{\langle x y \rangle} 
s_{xy} \left(\psi_x^\dagger \psi_y + \psi_y^\dagger \psi_x \right) + 
m \sum_x s_x \psi_x^{i \dagger} \psi_x^i. 
\end{equation}
Here $t$ is a hopping parameter, and $s_x$, $s_{xy}$ are sign-factors 
associated with the points $x$ and with the links connecting neighboring 
lattice points $x$ and $y$, respectively. The site factor is given by 
$s_x = (-1)^{x_1 + \dots + x_d}$. For the links in the 1-direction $s_{xy} = 1$, for 
those in the 2-direction $s_{xy} = (-1)^{x_1}$, and for the links in the 
$k$-direction $s_{xy} = (-1)^{x_1 + \dots + x_{k-1}}$. The sign-factors associated 
with the links represent a fixed $\Z(2)$ background ``gauge'' field, with a 
$\pi$-flux on each plaquette, which replaces the $\gamma$-matrices of standard
Dirac fermions. The free staggered fermion Hamiltonian of eq.(\ref{Hfree}) is
even simpler than the one of the Hubbard model, which also incorporates spin as 
well as an on-site fermion repulsion. Since there are implementations of quantum
simulators for the Hubbard model using ultracold matter in optical lattices, 
it is straightforward to construct quantum simulators for staggered fermions as
well. However, it is more non-trivial to quantum simulate dynamical gauge 
fields.

Before we approach this problem, let us first ask how staggered lattice
fermions propagate in a classical electromagnetic background field. In order to
address this question, we must first translate the continuum gauge field 
$A_k(x)$ to a lattice gauge field. Since $A_k(x)$ is a vector, its lattice 
variant is naturally associated not with a single lattice site $x$, but with the
directed link connecting neighboring sites $x$ and $y = x + \hat k$. When a 
gauge theory is regularized on the lattice, it is vital to maintain its 
invariance under gauge transformations. Since a continuum gauge transformation 
$A_k(x)' = A_k(x) - \partial_k \alpha(x)$ involves a derivative, and thus the
infinitesimal neighborhood of a point $x$, it is natural to integrate the
continuum gauge field over the link connecting $x$ and $y$. In fact, one can 
follow Wilson and construct the parallel transporter 
\begin{equation}
\label{WilsonU}
U_{xy} = \exp\left[i e \int_{x_k}^{x_k+a} dx_k \ A_k(x)\right] \in
U(1),  
\end{equation}
which takes values in the Abelian $U(1)$ gauge group of QED. Under gauge
transformations of the continuum gauge field $A_k(x)$, the parallel transporter 
transforms as
\begin{eqnarray}
U_{xy}'&=&\exp\left[i e \int_{x_k}^{x_k+a} dx_k \ A_k'(x)\right] =
\exp\left[i e \int_{x_k}^{x_k+a} dx_k \ 
\{A_k(x) - \partial_k \alpha(x)\}\right] \nonumber \\
&=&\exp\left[i e \left\{\int_{x_k}^{x_k+a} dx_k \ 
A_k(x) + \alpha(x) - \alpha(y)\right\}\right] = 
\Omega_x U_{xy} \Omega_y^\dagger,
\end{eqnarray}
where the lattice gauge transformation is given by
$\Omega_x = \exp\left[i e \alpha(x)\right] \in U(1)$. Just like the wave 
function of eq.(\ref{wavef}), the staggered fermion field operators transform as
$\psi_x' = \Omega_x \psi_x$, and the Hamiltonian of staggered fermions hopping 
in the background of a classical electromagnetic field takes the form
\begin{equation}
H = - t \sum_{\langle x y \rangle} s_{xy} \left(\psi_x^\dagger U_{xy} \psi_y + 
\psi_y^\dagger U_{xy}^\dagger \psi_x \right) + 
m \sum_x s_x \psi_x^{i \dagger} \psi_x^i. 
\end{equation}

\subsection{Lattice Quantum Electrodynamics}

In the previous subsections we have considered the quantum physics of charged
particles in an external classical electromagnetic field. Now we want to treat 
the field itself as a quantum object. Let us consider the QED Lagrange density
\begin{equation}
{\cal L}_{\text{QED}} = \overline\psi (i \gamma^\mu D_\mu - m) \psi - 
\frac{1}{4} F^{\mu\nu} F_{\mu\nu}.
\end{equation}
Here $\psi(x)$ and $\overline\psi(x) = \psi(x)^\dagger \gamma^0$ (with $x$ now
denoting a point in $(3+1)$-d space-time) are 4-component Dirac spinors 
describing electrons and positrons and $\gamma^\mu$ are the Dirac matrices. The 
covariant derivative $D_\mu = \partial_\mu + i e A_\mu(x)$ contains the 4-vector 
potential $A_\mu(x) = (\Phi(x),- \vec A(x))$, whose field strength tensor 
$F_{\mu\nu} = \partial_\mu A_\nu(x) - \partial_\nu A_\mu(x)$ contains the electric 
and magnetic fields $\vec E(x)$ and $\vec B(x)$. The $U(1)$ gauge symmetry 
reflects itself in the invariance of ${\cal L}_{\text{QED}}$ under gauge 
transformations
\begin{equation}
\psi(x)' = \exp[i e \alpha(x)] \psi(x), \quad
\overline \psi(x)' = \overline \psi(x) \exp[- i e \alpha(x)], \quad
A_\mu(x)' = A_\mu(x) - \partial_\mu \alpha(x).
\end{equation}
At the classical level, the energy density of the electromagnetic field is 
described by the Hamilton density 
${\cal H} = \frac{1}{2}(\vec E^2 + \vec B^2)$. In the absence of charges, the 
Hamilton operator of quantum electrodynamics is given by 
$H = \int d^3x \ {\cal H}$, however, $\vec E$ and $\vec B$ are now operators 
acting in a Hilbert space. In particular, the electric field 
operator is identified as the canonically conjugate momentum to the vector 
potential, which is given by $E_k(\vec x) = - i \partial/\partial A_k(\vec x)$.
Although in eq.(\ref{WilsonU}) we have constructed the parallel transporters 
from an underlying continuum vector potential $A_k(x)$, in lattice gauge theory 
$U_{xy}$ is the truly fundamental degree of freedom, from which a vector 
potential $A_{xy}$ could be derived as $U_{xy} = \exp\left(i a e A_{xy}\right)$.
The electric field operator associated with a link (in which we absorb the
factor $a e$ for convenience) is then given by 
$E_{xy} = - i \partial/\partial (a e A_{xy})$. This gives rise to the commutation
relations
\begin{equation}
\label{commuteEU}
[E_{xy},U_{x'y'}] = \delta_{xx'} \delta_{yy'} U_{xy}, \quad 
[E_{xy},U_{x'y'}^\dagger] = - \delta_{xx'} \delta_{yy'} U_{xy}^\dagger,
\end{equation}
i.e.\ operators associated with separate links commute with each other. On a 
lattice, the operator $\vec B$ is represented by the plaquette product 
$U_\Box =  U_{wx} U_{xy} U_{zy}^\dagger U_{wz}^\dagger$, such that the Hamiltonian of 
staggered fermions interacting with a dynamical electromagnetic quantum gauge 
field is given by
\begin{equation}
\label{Hamiltonian}
H_{\text{QED}} = 
- t \sum_{\langle x y \rangle} s_{xy} \left(\psi_x^\dagger U_{xy} \psi_y + 
\psi_y^\dagger U_{xy}^\dagger \psi_x \right) + 
m \sum_x s_x \psi_x^\dagger \psi_x + 
\frac{e^2}{2} \sum_{\langle x y \rangle} E_{xy}^2 - \frac{1}{4 e^2}
\sum_\Box \left(U_\Box + U_\Box^\dagger\right).
\end{equation}

One may wonder why, in the Hamiltonian formulation of quantum electrodynamics, 
we are not encountering the scalar potential $\Phi$. In the Lagrangian
formulation, which is used in Feynman's path integral quantization, $\Phi$ 
indeed arises as a Lagrange multiplier field that enforces the Gauss law 
$\vec \nabla \cdot \vec E = \rho$. In the Hamiltonian formulation, on the 
other hand, Gauss' law can not be implemented as an operator identity. In
particular, the operator $\vec \nabla \cdot \vec E$ does not vanish
in the absence of charges.  The lattice variant of the divergence of the 
electric field is given by the operator $\sum_k  (E_{x,x+\hat k} - E_{x-\hat k,x})$, 
while the charge density is $- \psi_x^\dagger \psi_x$. The Hermitean operator 
that generates an infinitesimal gauge transformation at the site $x$ and 
commutes with the Hamiltonian takes the form
\begin{equation}
\label{Gauss}
G_x = \psi_x^\dagger \psi_x + 
\sum_k \left(E_{x,x+\hat k} - E_{x-\hat k,x}\right), \quad [H,G_x] = 0.
\end{equation}
A general gauge transformation $\Omega_x$ is represented by the unitary
operator $V = \prod_x \exp(i e \alpha_x G_x)$ and acts as
\begin{equation}
\label{gaugetrafo}
V \psi_x V^\dagger = \Omega_x \psi_x, \quad 
V \psi_x^\dagger V^\dagger = \psi_x^\dagger \Omega_x^\dagger, \quad 
V U_{xy} V^\dagger = \Omega_x U_{xy} \Omega_y^\dagger.
\end{equation}
While the Hamiltonian is gauge invariant (it commutes with $G_x$ for all $x$),
most of its eigenstates are gauge variant. According to the Gauss law, those 
states do not belong to the physical Hilbert space, which contains only the
gauge invariant states $|\Psi\rangle$ that obey $G_x|\Psi\rangle = 0$. Despite
this restriction, the Hilbert space of Wilson's lattice gauge theory is
infinite-dimensional, even for a single link. This is because Wilson's parallel
transporter $U_{xy} \in U(1)$ is a continuous classical variable. Actually, each
link variable is analogous to a ``particle'' moving on the group manifold, which
is a circle for the Abelian gauge group $U(1)$. Since a particle on a circle 
has an infinite-dimensional Hilbert space, the same is true for Wilson's
lattice gauge theory.

\subsection{Abelian Quantum Link Models}

Quantum link models represent a generalization of lattice gauge theory 
beyond Wilson's framework, in which Wilson's continuous classical parallel 
transporters $U_{xy}$ are replaced by discrete quantum degrees of freedom,
called quantum links. Remarkably, although quantum links are discrete, they
exactly implement continuous gauge symmetries. $U(1)$ and $SU(2)$ quantum link 
models were first constructed by Horn in 1981, and further investigated by 
Orland and Rohrlich under the name of gauge magnets. In \cite{Cha97} quantum 
link models were introduced as an alternative non-perturbative regularization 
of gauge field theories. In a $U(1)$ quantum link model, the quantum link 
operator as well as the electric field operator are defined in terms of a 
quantum spin operator $\vec S_{xy}$ associated with a given link $xy$
\begin{equation}
U_{xy} = S_{xy}^1 + i S_{xy}^2 = S_{xy}^+, \quad 
U_{xy}^\dagger = S_{xy}^1 - i S_{xy}^2 = S_{xy}^-, \quad E_{xy} = S_{xy}^3.
\end{equation}
The quantum link operators $U_{xy}$ and $U_{xy}^\dagger$ act as raising and 
lowering operators of the electric flux $E_{xy}$. As a consequence of the spin 
commutation relations, 
$[S_{xy}^a,S_{x'y'}^b] = i \delta_{xx'} \delta_{yy'} \epsilon_{abc} S_{xy}^c$, 
$U_{xy}$, $U_{xy}^\dagger$, and $E_{xy}$ obey the same
commutation relations eq.(\ref{commuteEU}) as the corresponding objects in 
Wilson's lattice gauge theory. This implies that the Hamiltonian of 
eq.(\ref{Hamiltonian}), the gauge generator of eq.(\ref{Gauss}), and the gauge
transformations of eq.(\ref{gaugetrafo}) all maintain their previous form.
However, in contrast to Wilson's theory, the Hilbert space of a quantum link 
model is finite-dimensional. When one chooses spin $S$ on each link, the link
Hilbert space is $(2S + 1)$-dimensional. In particular, when one chooses 
$S = \frac{1}{2}$, there are just two discrete states per link, representing two
possible values $E_{xy} = \pm \frac{1}{2}$ of the electric flux, and still the
system has an exact continuous $U(1)$ gauge symmetry. While in Wilson's theory
$[U_{xy},U_{x'y'}^\dagger] = 0$, in a $U(1)$ quantum link model
$[U_{xy},U_{x'y'}^\dagger] = 2 \delta_{xx'} \delta_{yy'} E_{xy}$. Remarkably, this 
modification does not compromise gauge invariance or other symmetries of the
theory.

Thanks to their discrete nature, quantum link models can be embodied by the 
quantum states of ultracold atoms. Due to their relation to quantum spins, 
quantum links can be represented by Schwinger bosons, 
$[b_x,b_y^\dagger] = \delta_{xy}$, 
$[b_x,b_y] = [b_x^\dagger,b_y^\dagger] = 0$,
\begin{equation}
U_{xy} = b_x b_y^\dagger, \quad U_{xy}^\dagger = b_y b_x^\dagger, \quad
E_{xy} = \frac{1}{2}(b_x^\dagger b_x - b_y^\dagger b_y),
\end{equation}
The spin $S$ determines the total number of bosons 
$b_x^\dagger b_x + b_y^\dagger b_y = 2S$ on each link.

\subsection{The $(2+1)$-d $U(1)$ Quantum Link Model}

In order to get a flavor of its rich physics, let us consider the $(2+1)$-d
$U(1)$ quantum link model with $S = \frac{1}{2}$. This model has also been 
investigated in the context of quantum spin liquids \cite{Her04}. We consider 
the model with a plaquette coupling $J$ and a Rokhsar-Kivelson coupling 
$\lambda$
\begin{equation}
H = - J \sum_{\Box} \left[U_\Box + U_\Box^\dagger - 
\lambda (U_\Box + U_\Box^\dagger)^2\right].
\end{equation}
The gauge invariant flux states that obey the Gauss law at a site $x$ are 
shown in Figure 1a. 
\begin{figure}[b]
\centering
\includegraphics[scale=0.6]{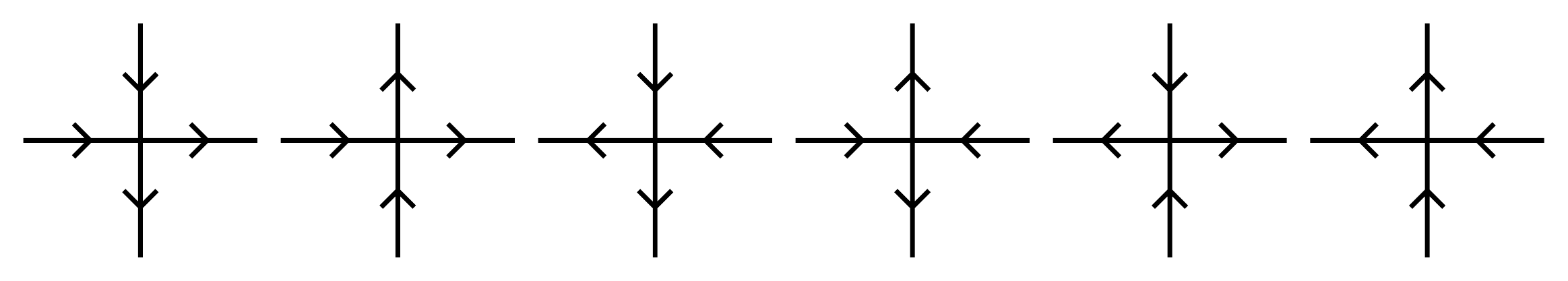}
\includegraphics[scale=0.75]{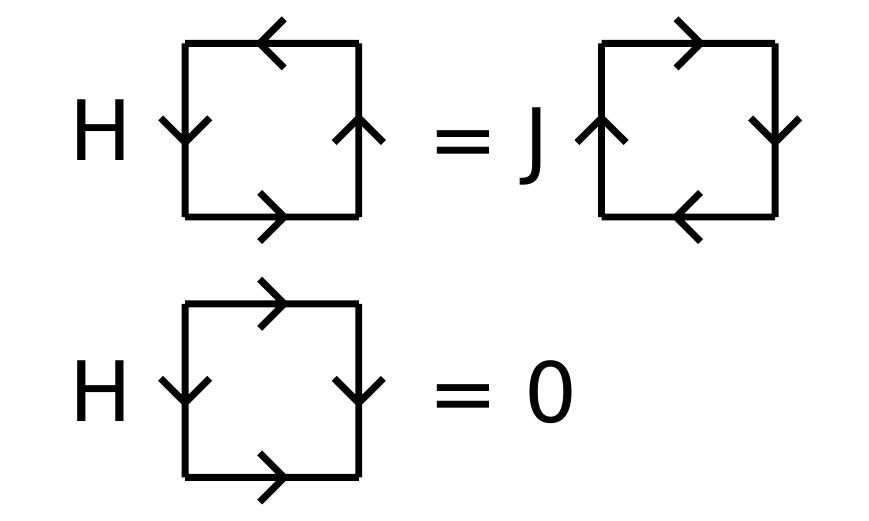}
\includegraphics[scale=0.6]{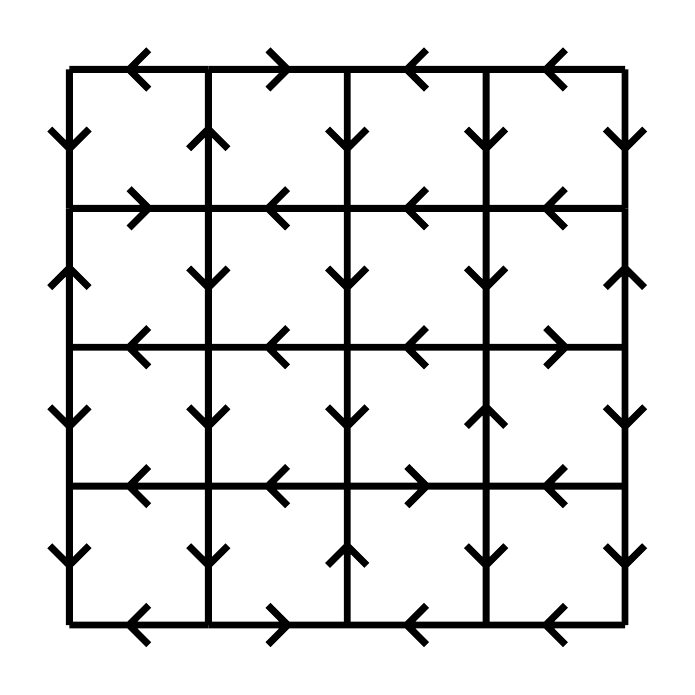}
\caption{Top: a) The six flux states that satisfy the Gauss law in the 
$(2+1)$-d $U(1)$ quantum link model with $S = \frac{1}{2}$ on each link. Bottom 
left: b) The $J$-term in the Hamiltonian reverses the direction of a closed loop
of flux (a flippable plaquette). It annihilates all other flux states. Bottom 
right: c) A typical flux configuration on a $4 \times 4$ lattice with periodic 
boundary conditions. The fluxes obey the local Gauss law at every site $x$.}
\end{figure}
As illustrated in Figure 1b, the $J$-term flips a loop of electric 
flux, winding around an elementary plaquette, and annihilates non-flippable 
plaquette states, while the Rokhsar-Kivelson term, proportional to $\lambda$, 
counts flippable plaquettes. The corresponding phase diagram is sketched in 
Fig.2a.
\begin{figure}[t]
\includegraphics[width=0.47\textwidth]{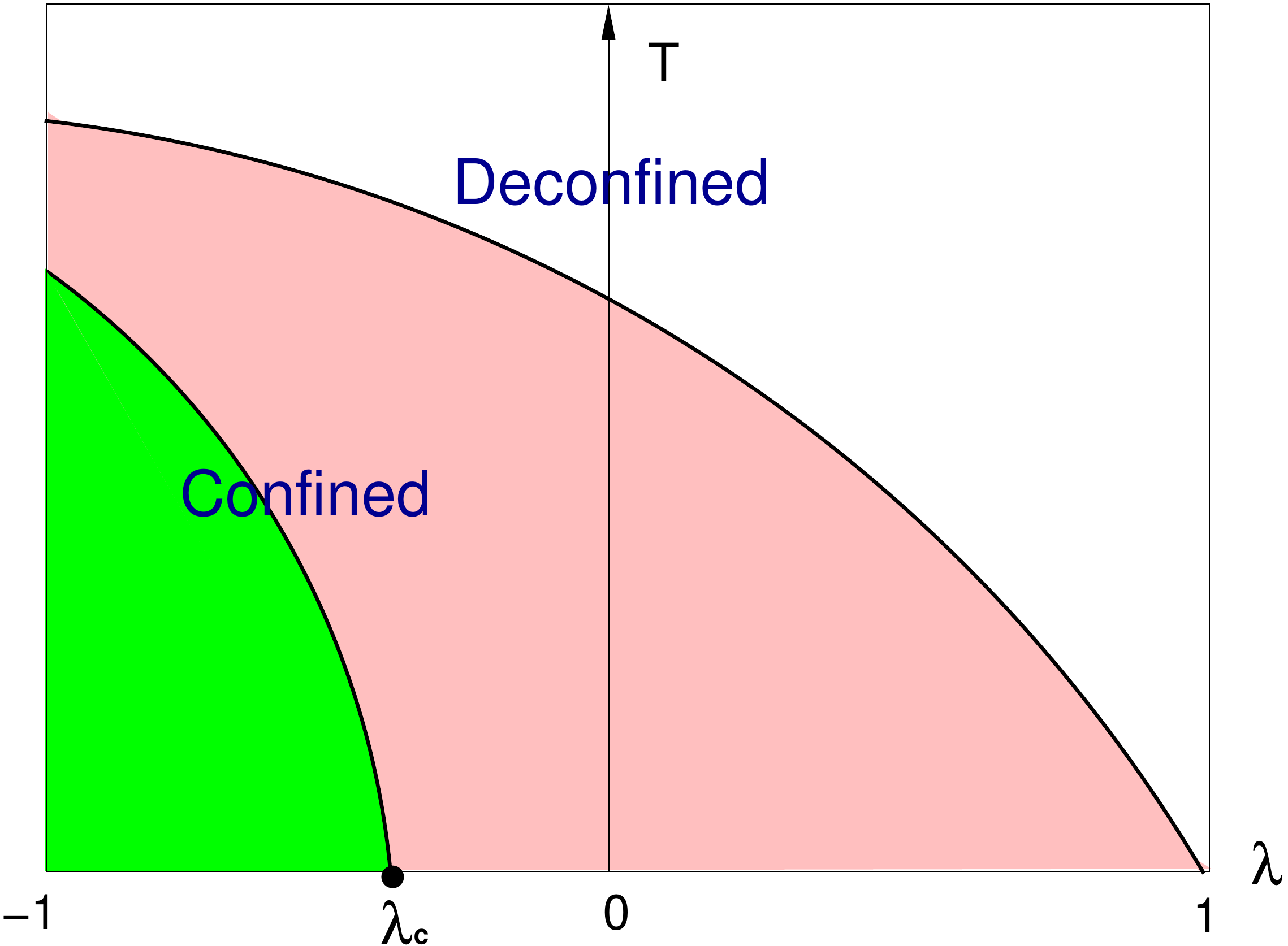}
\includegraphics[width=0.53\textwidth]{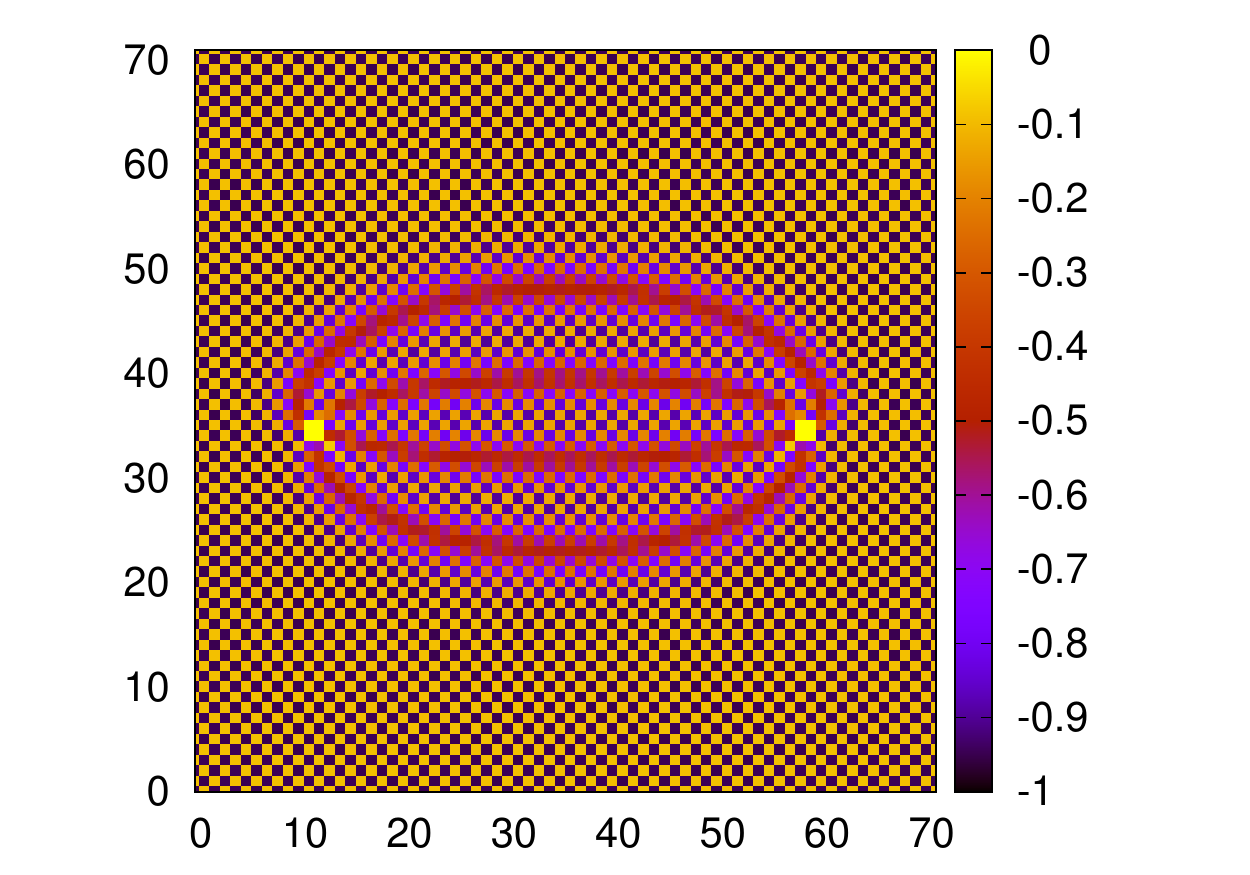}
\caption{\textit{Left: a) Schematic sketch of the $\lambda$-$T$ 
phase diagram. Right: b) Energy density $-J \langle U_\Box + U_\Box^\dagger\rangle$
in the presence of two charges $\pm 2$ for $\lambda = 0$ and $T = 0$.}}
\end{figure}
At zero temperature, the model is confining for $\lambda < 1$, while at finite 
temperature $T$ it has a deconfinement phase transition. At $\lambda_c$ there 
is a quantum phase transition which separates two confined phases with 
spontaneously broken translation symmetry. The phase at $\lambda < \lambda_c$ 
has, in addition, a spontaneously broken charge conjugation symmetry. The two 
phases are similar to the columnar and plaquette ordered valence bond solid 
phases in a quantum dimer model \cite{Moe02}. In fact, the quantum dimer model
has the same Hamiltonian as the quantum link model, but replaces the usual Gauss
law by a staggered background of static charges $\pm 1$. It turns out that the
quantum phase transition masquerades as a deconfined quantum critical point, at
which an approximate spontaneously broken global $SO(2)$ symmetry with an
almost massless pseudo-Goldstone boson emerges dynamically \cite{Ban13a}. 
However, since the Goldstone boson is not exactly massless, it cannot be 
interpreted as a dual photon, and the theory remains confining at the quantum 
phase transition, albeit with a rather small string tension. An unbreakable 
confining string has an energy proportional to its length, with the string
tension being the proportionality factor. In the $(2+1)$-d $U(1)$ quantum link 
model, the confining strings display unusual features. As illustrated in Figure 
2b, the strings connecting two 
external charges $\pm 2$ separate into four mutually repelling strands, which
each carry fractional electric flux $\frac{1}{2}$. The interior of the strands 
consists of the bulk phase that is stable on the other side of the phase 
transition. For integer spin $S$ on each link, one expects a less exotic 
confining dynamics. Once the $(2+1)$-d $U(1)$ quantum link model is realized in 
ultracold matter experiments, its dynamics in real-time will become accessible 
to quantum simulation. Since the model can be efficiently simulated in Euclidean
time, experimental realizations of quantum simulators can be bench-marked 
accurately. For sufficiently large $S$, and perhaps even for $S = \frac{1}{2}$,
the $(3+1)$-d $U(1)$ quantum link model exists in a Coulomb phase \cite{Cha99}, 
which is interpreted as a spin liquid in condensed matter parlance \cite{Her04}.

\section{Quantum Simulators for Abelian Lattice Gauge Theories}

In this section we discuss constructions of quantum simulators for dynamical
Abelian gauge theories, with and without matter fields, using both digital and
analog simulation concepts.
 
\subsection{Digital Quantum Simulators for $\Z(2)$ and $U(1)$ Gauge Theories}

A digital quantum simulator is a precisely controllable many-body system, which
can be programmed to execute a prescribed sequence of quantum gate operations.
The state of the simulated system is encoded as quantum information, and its
dynamics is represented by a sequence of quantum gates, following a stroboscopic
Trotter decomposition \cite{Llo96}. The feasibility of universal digital quantum
simulation with trapped ions has been demonstrated with 6 qubits and up to
100 gate operations in \cite{Lan11}, where multi-spin interactions have also
been implemented. While trapped ions offer a large degree of control of 
complicated interactions, these systems are still limited to a relatively small 
number of qubits and are hence currently not easily scalable. Utilizing 
engineered dissipative processes, entangled 4-qubit states, representing a 
single plaquette of Kitaev's toric code, have been quantum simulated with 
trapped ions \cite{Bar11}. The toric code is equivalent to a $\Z(2)$ quantum 
link model on a 2-d quadratic spatial lattice with the Hamiltonian 
\begin{equation}
H = - J \sum_\Box U_\Box - G \sum_x V_x, \quad
V_x = \exp\left[i \pi \sum_k (S^3_{x,x+\hat k} - S^3_{x-\hat k,x})\right],
\end{equation}
As in the $(2+1)$-d $U(1)$ quantum link model, there is a spin $\frac{1}{2}$, 
$\vec S_{xy}$, on each link. However, the quantum link operator is now given by
$U_{xy} = S_{xy}^1$, such that the plaquette term 
$U_\Box = S_{wx}^1 S_{xy}^1 S_{yz}^1 S_{zw}^1$ is invariant under discrete 
$\Z(2)$ gauge transformations $\Omega_x = \pm 1$. The $G$-term with $G > 0$ 
punishes violations of the $\Z(2)$ Gauss law $V_x |\Psi\rangle = |\Psi\rangle$.

Optical lattices arise from the interference of counter-propagating laser 
beams. The lattice can be 1-, 2-, or 3-dimensional, with about 100 lattice
sites per direction, and with a lattice spacing of a fraction of a $\mu$m.
The resulting periodic structure of light can be loaded with up to about $10^5$ 
atoms from an ultracold Bose-Einstein condensate, which then settle down in the 
corresponding
potential wells. By varying the depth of the optical potential, one controls
the tunneling rate of atoms between neighboring wells. In this way, the phase 
transition between a Mott insulator of localized bosonic atoms and a superfluid 
has been quantum simulated \cite{Gre02}. Using lasers one can excite atoms to 
high-lying Rydberg states. Rydberg atoms are large objects with strong 
long-range dipole-dipole interactions, which can give rise to collective 
interactions of several atoms. Rydberg atoms in an optical lattice with a 
large lattice spacing can be addressed individually by external lasers. A
Rydberg gate allows to entangle a number of atoms with a single control atom 
\cite{Mue09}. This is the basis of theoretical constructions of digital 
quantum simulators for $U(1)$ quantum link models \cite{Wei10,Tag13}, which use 
control atoms at lattice sites to ensure Gauss' law, as well as at plaquette 
centers to facilitate the flip of electric flux loops wrapping around flippable
plaquettes. The ensemble Rydberg atoms represent qubits that reside at the link 
centers. By optical pumping, one can engineer dissipation that leads the system
into the ground state, e.g.\ at the Rokhsar-Kivelson point $\lambda = 1$. By 
further adiabatic evolution, one can then reach the ground state at any desired 
point in parameter space, and investigate, e.g., the dynamics of confining 
strings or the nature of quantum phase transitions.

\subsection{Analog Quantum Simulators for $U(1)$ Gauge Theories}

In an analog quantum simulator the time evolution proceeds continuously, 
rather than through a discrete sequence of quantum gates. These devices are
usually limited to simpler interactions, but they are more easily scalable to
large system sizes. The first quantum simulator for a gauge theory was 
proposed in \cite{Bue05}. It addresses the physics of the $U(1)$ quantum link
model with $S = \frac{1}{2}$ using ultracold atoms in an optical lattice. The
quantum links are embodied by hard-core bosons placed on the link centers. Their
ring-exchange plaquette term is induced by a Raman transition that couples the
bosons to a ``molecular'' two particle state localized at the plaquette center.

Zohar and Reznik have constructed a quantum simulator for Wilson's compact 
$U(1)$ pure gauge theory \cite{Zoh11}. In order to represent the continuously 
varying complex phase of $U_{xy} \in U(1)$, which gives rise to an 
infinite-dimensional Hilbert space per link, they have proposed to place a 
Bose-Einstein condensate on each link. In collaboration with Cirac
\cite{Zoh12}, they have simplified this construction by restricting the number
of bosons per link to $2S$, which truncates the Wilson theory to a $U(1)$ 
quantum link model with just a $(2S+1)$-dimensional link Hilbert space.
Compact $(3+1)$-d $U(1)$ gauge theory describes magnetic monopoles 
interacting with photons. When the monopoles condense, the theory is confining
\cite{Pol77}, and a quantum simulator can thus address the corresponding
non-perturbative effects. The confined phase at strong coupling is separated 
from a weakly coupled Coulomb phase by a first order phase transition. In the 
Coulomb phase, the monopoles have a mass at the cut-off scale, and the 
continuum limit just describes free photons. As was argued in \cite{Cha99},
a quantum link model with small $S$ is sufficient to reach the continuum limit, 
and there is no need to approach the Wilson theory by increasing $S$ any 
further.

\begin{figure}[b]
\begin{center}
\includegraphics[width=0.45\textwidth]{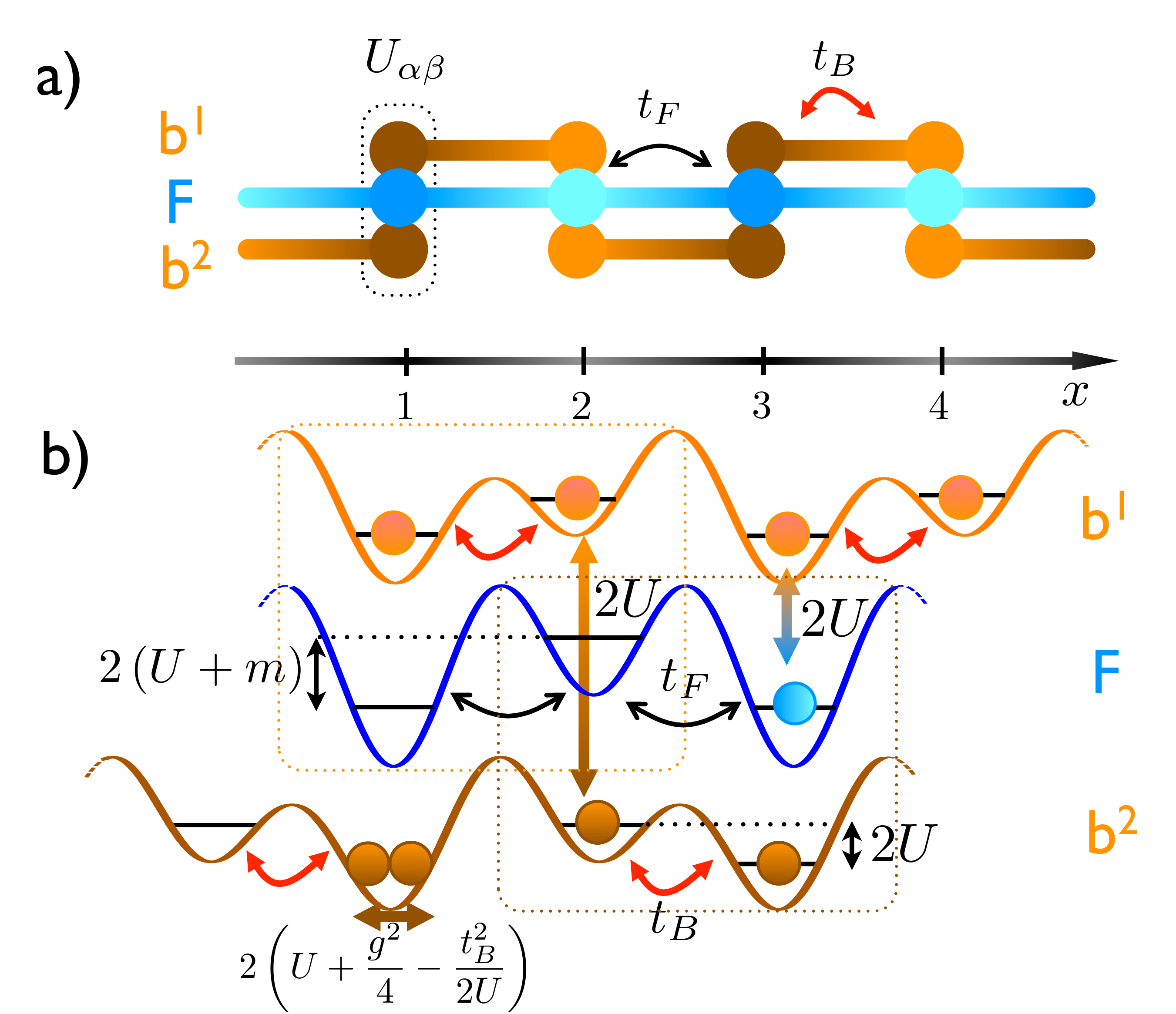} \hspace{1cm}
\includegraphics[width=0.45\textwidth]{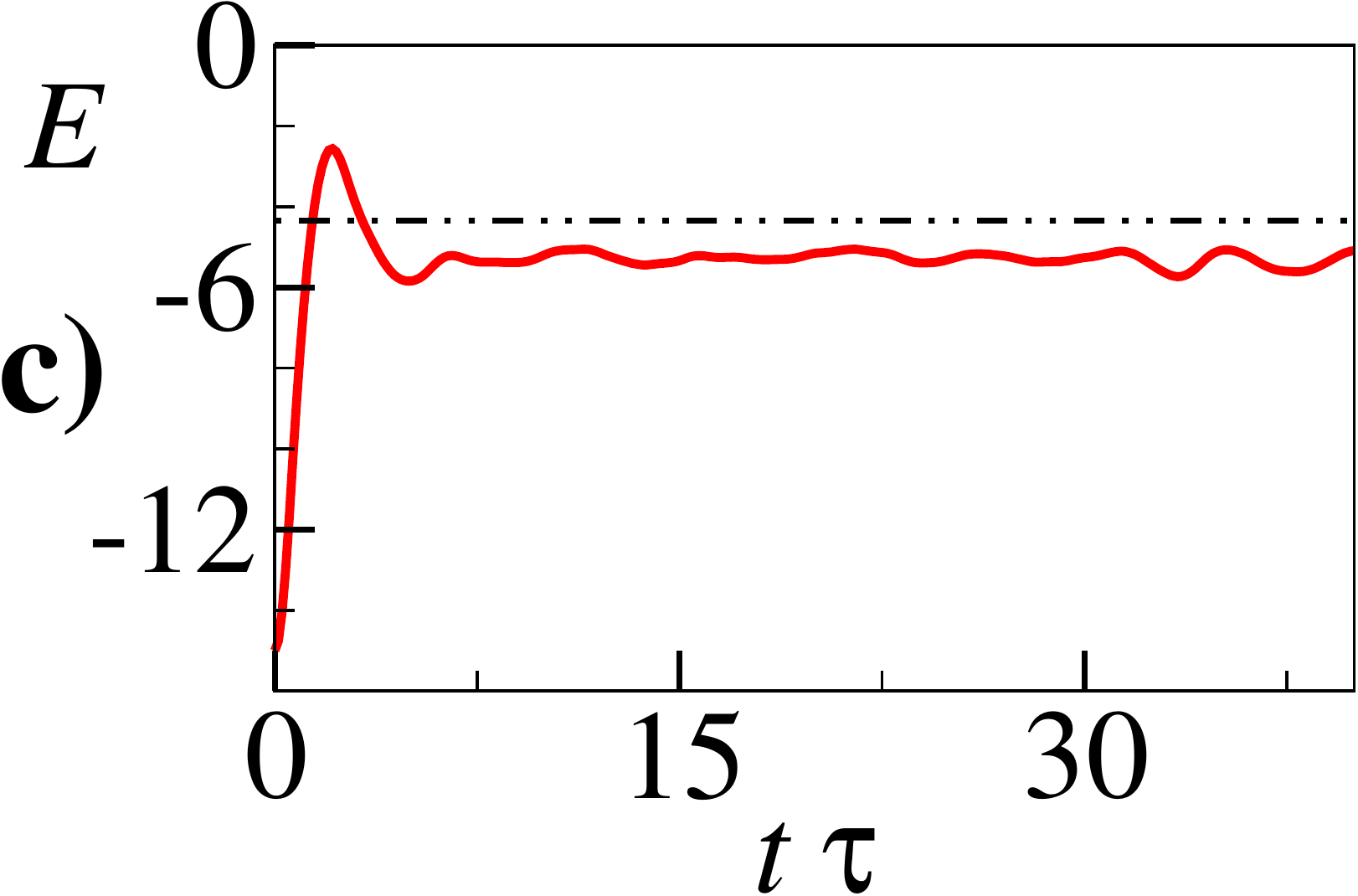}
\caption{\it a) Quantum simulator of a $(1+1)$-d $U(1)$ quantum link model with 
fermionic matter using a Bose-Fermi mixture of cold atoms hopping in a 3-strand 
optical lattice. b) Mass $m$, hopping $t$, and interaction $U$ parameters
are encoded in an optical superlattice. Energy conservation favors the 
simultaneous hopping of fermions and bosons, thus ensuring gauge invariance at
low energies. c) Dynamical string breaking in real time $\tau$ (in
units of the hopping parameter $t$): the string's energy is converted into the 
mass of a dynamically created charge-anti-charge pair. In this process, the 
large negative electric flux $E$ of the string quickly relaxes to its vacuum
value \cite{Ban12}.}
\end{center}
\end{figure}
The physics becomes more interesting, and much more difficult to simulate on a 
classical computer, when dynamical fermions are added to the gauge field.
Kapit and Mueller have proposed to use fermionic atoms hopping on a 2-d
graphene-like optical honeycomb lattice, to engineer relativistic fermions at
the corresponding Dirac cones \cite{Kap11}. In their construction, three bosonic
species represent the components of the non-compact vector potential $A_\mu(x)$ 
of the $(2+1)$-d continuum theory. Since the theory was gauge-fixed in the
continuum before it was discretized on a lattice, it is unclear whether, in this
case, light gauge degrees of freedom emerge naturally.

In \cite{Ban12} a Bose-Fermi mixture was proposed to quantum simulate $U(1)$ 
quantum links, embodied by $2S$ bosons per link, coupled to dynamical staggered 
fermions. While the construction works in any dimension, it is particularly 
simple in 1-d, where it uses the 3-strand optical superlattice illustrated in 
Figure 3a. The Hamiltonian of eq.(\ref{Hamiltonian}) (without the 
plaquette term) then arises in second order perturbation theory from a 
microscopic Hubbard-type Hamiltonian
\begin{equation}
\widetilde H = \sum_x h^B_{x,x+1} + \sum_x h^F_{x,x+1} + m \sum_x (-1)^x n_x^F +
U \sum_x \widetilde G_x^2,
\end{equation}
which punishes gauge-variant states by a large energy penalty $U$. Here
\begin{equation}
\widetilde G_x = n_x^F + n^B_x - 2S + \frac{1}{2}\left[(-1)^x - 1 \right],
\end{equation}
is the infinitesimal generator of gauge transformations, with $n_x^F$ and 
$n^B_x$ counting the number of dynamical fermions and Schwinger bosons at the
site $x$. The bosons that represent the $U(1)$ gauge field are confined to a
given link, and just hop between the two ends $x$ and $x+1$ of that link,
according to the hopping term 
$h^B_{x,x+1} = - t_B (b^\dagger_{x+1} b_x + b^\dagger_x b_{x+1})$. The number 
$b^\dagger_x b_x + b^\dagger_{x+1} b_{x+1} = 2S$ of bosonic atoms is conserved 
locally on each link. In addition, one has spinless fermionic atoms at 
half-filling, which can hop between neighboring sites throughout the entire 
system, based on the hopping term 
$h^F_{x,x+1} = - t_F (\psi_{x+1}^\dagger \psi_x + \psi_x^\dagger \psi_{x+1})$. Due to
the large energy penalty $U$, a fermion hopping from $x$ to $x+1$ is necessarily
accompanied by a boson hopping the other way, thus ensuring gauge invariance in
the low-energy sector by energy conservation (c.f.\ Figure 3b). 
This simple model can be used to study the dynamics of string breaking by 
charge-anti-charge pair creation in real time, a process which also arises in 
QCD. As illustrated schematically in Figure 4, an external static 
quark-anti-quark pair $\bar Q Q$ is connected by a confining electric flux 
string (Figure 4b), which manifests itself by a large value of the electric 
flux. For sufficiently small fermion mass, the potential energy stored in the 
string is converted into kinetic energy by fermion hopping, which amounts to 
the creation of a dynamical quark-anti-quark pair $q \bar q$ (Figures 4c,d).
In the process of string breaking, an external static anti-quark $\bar Q$ pairs 
up with a dynamical quark to form a $\bar Q q$ meson (c.f.\ Figures 4e).
\begin{figure}
\centering
\setlength{\unitlength}{3pt}
\begin{picture}(80,75)(-5,-10)

\multiput(0,0)(0,15){5}{\line(1,0){70}}

\multiput(0,60)(20,0){4}{\color{red}\circle*{1.5}}
\multiput(10,60)(20,0){4}{\put(0,-1){\line(0,1){2}}}
\put(-5,59){$a)$}
\put(-1,63){$\frac{1}{2}$}
\put(67,63){$-\frac{1}{2}$}

\multiput(0,45)(20,0){4}{\color{red}\circle*{1.5}}
\multiput(10,45)(20,0){4}{\put(0,-1){\line(0,1){2}}}
\multiput(3.0,45)(10,0){7}{\thicklines \color{blue}\vector(-1,0){0}}
\multiput(5.0,45)(10,0){7}{\thicklines \color{blue}\vector(-1,0){0}}
\put(-5,44){$b)$}
\put(-1.5,48){$\bar Q$}
\put(68.5,48){$Q$}

\multiput(0,30)(20,0){2}{\color{red}\circle*{1.5}}
\put(30,30){\color{red}\circle*{1.5}}
\put(60,30){\color{red}\circle*{1.5}}
\put(10,30){\put(0,-1){\line(0,1){2}}}
\put(40,30){\put(0,-1){\line(0,1){2}}}
\multiput(50,30)(20,0){2}{\put(0,-1){\line(0,1){2}}}
\multiput(3.0,30)(10,0){3}{\thicklines \color{blue}\vector(-1,0){0}}
\multiput(5.0,30)(10,0){3}{\thicklines \color{blue}\vector(-1,0){0}}
\multiput(43.0,30)(10,0){3}{\thicklines \color{blue}\vector(-1,0){0}}
\multiput(45.0,30)(10,0){3}{\thicklines \color{blue}\vector(-1,0){0}}
\put(-5,29){$c)$}
\put(-1.5,33){$\bar Q$}
\put(29,33){$q$}
\put(39,33){$\bar q$}
\put(68.5,33){$Q$}

\put(0,15){\color{red}\circle*{1.5}}
\multiput(10,15)(20,0){3}{\color{red}\circle*{1.5}}
\multiput(20,15)(20,0){3}{\put(0,-1){\line(0,1){2}}}
\put(70,15){\put(0,-1){\line(0,1){2}}}
\multiput(3.0,15)(20,0){4}{\thicklines \color{blue}\vector(-1,0){0}}
\multiput(5.0,15)(20,0){4}{\thicklines \color{blue}\vector(-1,0){0}}
\put(-5,14){$d)$}
\put(-1.5,18){$\bar Q$}
\put(9,18){$q$}
\put(19,18){$\bar q$}
\put(29,18){$q$}
\put(39,18){$\bar q$}
\put(49,18){$q$}
\put(59,18){$\bar q$}
\put(68.5,18){$Q$}

\multiput(0,0)(20,0){3}{\color{red}\circle*{1.5}}
\put(10,0){\color{red}\circle*{1.5}}
\multiput(30,0)(20,0){3}{\put(0,-1){\line(0,1){2}}}
\put(60,0){\put(0,-1){\line(0,1){2}}}
\put(3.0,0){\thicklines \color{blue}\vector(-1,0){0}}
\put(5.0,0){\thicklines \color{blue}\vector(-1,0){0}}
\put(63.0,0){\thicklines \color{blue}\vector(-1,0){0}}
\put(65.0,0){\thicklines \color{blue}\vector(-1,0){0}}

\put(-5,-1){$e)$}
\put(-1.5,3){$\bar Q$}
\put(9,3){$q$}
\put(59,3){$\bar q$}
\put(68.5,3){$Q$}

\put(1,-3){\oval(2,2)[bl]}
\put(1,-4){\line(1,0){3}}
\put(4,-5){\oval(2,2)[tr]}
\put(6,-5){\oval(2,2)[tl]}
\put(6,-4){\line(1,0){3}}
\put(9,-3){\oval(2,2)[br]}
\put(-0.5,-9){$meson$}

\put(11,-3){\oval(2,2)[bl]}
\put(11,-4){\line(1,0){23}}
\put(34,-5){\oval(2,2)[tr]}
\put(36,-5){\oval(2,2)[tl]}
\put(36,-4){\line(1,0){23}}
\put(59,-3){\oval(2,2)[br]}
\put(28.5,-9){$vacuum$}

\put(61,-3){\oval(2,2)[bl]}
\put(61,-4){\line(1,0){3}}
\put(64,-5){\oval(2,2)[tr]}
\put(66,-5){\oval(2,2)[tl]}
\put(66,-4){\line(1,0){3}}
\put(69,-3){\oval(2,2)[br]}
\put(59.5,-9){$meson$}

\end{picture}
\caption{\it a) Half-filled vacuum of the $(1+1)$-d $U(1)$ quantum link model
for $S = 1$ with staggered fermions, b) string induced by a static external 
$\bar Q Q$ pair, c) broken string, d) evolution, e) final state with two mesons 
separated by vacuum. The dots denote sites occupied by a fermion and the 
double-arrows denote non-zero electric fluxes $E_{xy} = S_{xy}^3 = - 1$.}
\end{figure}
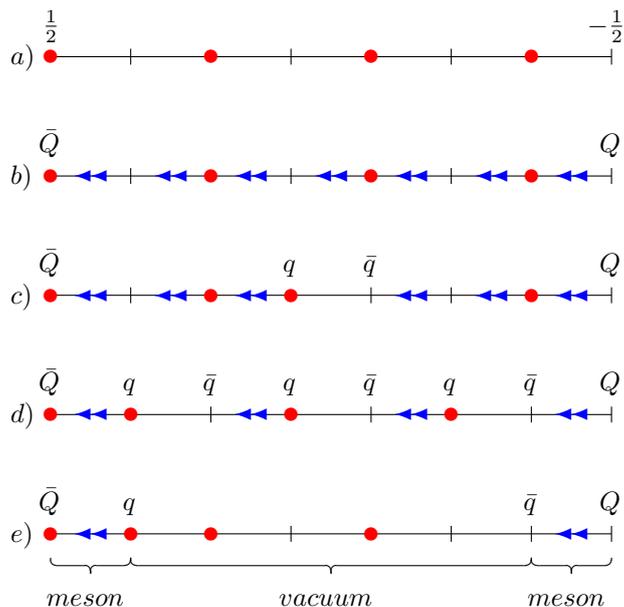
To illustrate what could be observed in a corresponding experiment, exact 
diagonalization results for the real-time evolution of the electric flux have 
been obtained for the quantum link model Hamiltonian with $S = 1$. Figure 3c 
shows string breaking, starting from the initial state shown in Figure 4b. The
large separation of the external static quark-anti-quark pair $\bar Q Q$ gives
rise to a large energy stored in the confining string, which is then converted
into the mass of two mesons. Indeed, the large negative electric flux initially 
stored in the string quickly approaches its vacuum value.

Quantum simulators for $U(1)$ gauge theories with fermionic matter have also 
been constructed in \cite{Zoh13}. In \cite{Zoh13b} Zohar, Cirac, and Reznik 
made the interesting observation that $U(1)$ gauge invariance can be protected
by angular momentum conservation in the interactions between the atoms
representing fermions and gauge fields. Gauge invariance is then naturally 
maintained, once the initial state obeys the Gauss law.

\section{Non-Abelian Gauge Theories}

Non-Abelian gauge theories play a central role in the Standard Model of particle
physics, which is a relativistic quantum field theory with the gauge group
$SU(3) \times SU(2)_L \times U(1)_Y$. By the Higgs mechanism, the gauge symmetry
gets broken to $SU(3) \times U(1)$, where $U(1)$ is the Abelian gauge group of 
electromagnetism and $SU(3)$ is the non-Abelian color gauge group controlling 
the strong interactions between quarks and gluons. Electromagnetism arises as a 
low-energy remnant of the electroweak symmetry $SU(2)_L \times U(1)_Y$, which 
governs the interactions of quarks, electrons, and neutrinos mediated by photons
as well as massive W and Z bosons. While electromagnetism
and the weak interactions can be understood in perturbation theory, the strong 
interactions require a non-perturbative treatment, at least at low energies. In
particular, the QCD interaction between quarks and gluons is so strong that they
do not exist as isolated particles, but are permanently confined inside protons,
neutrons, and other hadrons. QCD can be regularized non-perturbatively on a 
lattice. In this framework, many important questions in particle physics are 
being addressed very successfully from first principles, using Monte Carlo
simulations. Besides determining the hadron spectrum \cite{Dur08,Baz10,Aok11}, 
which is vital for confirming QCD as the correct theory of the strong 
interactions also at low energies, lattice QCD calculations are important for 
the correct extraction of the quark masses by comparison with experiments. 
Furthermore, lattice QCD is vital for understanding the structure of hadrons 
and for investigating the physics of quarks and gluons at high temperatures 
\cite{Aok06,Baz09}, which is explored in heavy ion collision experiments, e.g., 
at RHIC and at the LHC. Despite the many successes of lattice QCD, there are 
also significant challenges. Due to severe sign problems, dense systems of 
strongly interacting matter, such as the core of neutron stars, cannot be 
addressed with lattice QCD Monte Carlo simulations. For the same reason, the 
real-time evolution of strongly interacting systems remains beyond reach. These 
unsolved challenging problems are a strong motivation to formulate QCD as well 
as other non-Abelian gauge theories in a way that makes them accessible to 
atomic quantum simulation.

First, we discuss QCD in the continuum. In order to address non-perturbative 
effects, we then formulate the theory on the lattice, following Wilson. In order
to ease the construction of quantum simulators, we also introduce non-Abelian 
quantum link models. We then present several quantum simulator constructions, 
and discuss potential experimental investigations.

\subsection{Quantum Chromodynamics}

Before we formulate QCD on the lattice, in this subsection we review some 
basic features of QCD in the continuum, using the manifestly relativistically 
invariant Lagrangian formulation. The QCD Lagrangian takes the form
\begin{equation}
{\cal L}_{\text{QCD}} = 
\sum_{f,i} \overline \psi^{fi} (i \gamma^\mu D_{\mu ij} - m_f) \psi^{fj} - 
\frac{1}{2 g^2}\mbox{Tr} (G^{\mu\nu} G_{\mu\nu}).
\end{equation}
The Dirac spinors $\psi^{fi}(x)$ and 
$\overline\psi^{fi}(x) = {\psi^{fi}(x)}^\dagger \gamma^0$ describe quarks and 
anti-quarks of different flavors $f \in \{1,2,\dots,N_f\}$ and colors 
$i \in \{1,2,\dots,N\}$. The lightest quarks, which dominate ordinary matter, 
carry the flavors up and down. Protons consist of two u-quarks (each 
with electric charge $\frac{2}{3} e$) and one d-quark (of electric charge 
$- \frac{1}{3} e$), while neutrons consist of one u-quark and two d-quarks. In 
addition, protons and neutrons contain a fluctuating number of gluons and 
quark-anti-quark pairs. The non-Abelian vector potential,
$G_\mu(x) = i g G_\mu^a(x) T^a$, describing the gluons is constructed from 
real-valued fields $G_\mu^a(x)$ multiplying the $N^2 -1$ traceless Hermitean 
generators $T^a$ of $SU(N)$ --- the group of unitary $N \times N$ matrices with 
determinant 1. In the real world the number of colors is $N = 3$. For $N = 2$ 
the generators $T^a = \frac{1}{2} \sigma^a$ are given by the Pauli matrices, 
while for $N = 3$ they are given by the Gell-Mann matrices 
$T^a = \frac{1}{2} \lambda^a$. Here $g$ is the strong coupling constant, i.e.\ 
the non-Abelian analog of the elementary electric charge $e$. The non-Abelian 
covariant derivative takes the form
\begin{equation}
D_\mu = \partial_\mu + G_\mu(x) \ \Rightarrow \
D_{\mu ij} = \partial_\mu \delta_{ij} + G_{\mu ij}(x) =
\partial_\mu \delta_{ij} + i g G_\mu^a(x) T_{ij}^a,
\end{equation}
and the gluon field strength tensor is given by
\begin{equation}
G_{\mu\nu}(x) = \partial_\mu G_\nu(x) - \partial_\nu G_\mu(x) + [G_\mu(x),G_\nu(x)].
\end{equation}
Unlike photons, which are electrically neutral, gluons carry color charge. This
manifests itself in the non-Abelian commutator term in $G_{\mu\nu}(x)$, which is 
absent in QED. The QCD Lagrangian is invariant under color gauge transformations
$\Omega(x) \in SU(N)$ of the quark and gluon fields
\begin{eqnarray}
&&\psi^f(x)' = \Omega(x) \psi^f(x) \ \Rightarrow
\psi^{fi}(x)' = \Omega_{ij}(x) \psi^{fj}(x), \nonumber \\
&&G_\mu(x)' = \Omega(x) [G_\mu(x) + \partial_\mu] \Omega(x)^\dagger \ \Rightarrow \
G_{\mu\nu}(x)' = \Omega(x) G_{\mu\nu}(x) \Omega(x)^\dagger.
\end{eqnarray}
It should be pointed out that, unlike $F_{\mu\nu}$ in Abelian gauge theories,
the non-Abelian field strength $G_{\mu\nu}$ is not gauge invariant. The gluon 
field couples to the color index $i$ of the quark field $\psi^{fi}(x)$, but does
not distinguish between quarks of different flavors $f$, which differ only in 
their masses $m_f$.

Quarks and anti-quarks are distinguished by their baryon numbers 
$B = \pm \frac{1}{N}$. In the real world (with $N = 3$) three quarks form a 
baryon (e.g.\ a proton or neutron), while three anti-quarks from an anti-baryon
(e.g.\ an anti-proton or anti-neutron). Under the global $U(1)_B$ baryon number 
symmetry the quark fields transform as 
$\psi^{fi}(x)' = \exp(i \alpha) \psi^{fi}(x)$,
which leaves ${\cal L}_{\text{QCD}}$ invariant. In the absence of quark masses, 
i.e.\ for $m_f = 0$, the QCD Lagrangian has a global $SU(N_f)_L \times SU(N_f)_R$
chiral symmetry acting separately on the left- and right-handed quark and 
anti-quark fields. At low temperature, chiral symmetry is spontaneously broken 
to its vector subgroup $SU(N_f)_{L=R}$, known as isospin for $N_f = 2$. The 
order parameter for this symmetry breaking is the chiral condensate 
$\langle \overline \psi \psi \rangle = 
\langle 0|\sum_{f,i} \overline\psi^{f,i}(x) \psi^{f,i}(x)|0\rangle$. Here 
$|0\rangle$ is the QCD vacuum state, the lowest energy eigenstate in the sector 
with baryon number $B = 0$. According to the Goldstone theorem, the 
spontaneous breakdown of chiral symmetry gives rise to $N_f^2-1$ Goldstone 
bosons --- 3 pions in the $N_f = 2$ case. In the real world, the masses $m_u$ 
and $m_d$ of the up and down quarks are small, but non-zero, which turns the 
pions into light, but not exactly massless, pseudo-Goldstone bosons. Besides 
the pions, the QCD spectrum contains other mesons (states with baryon
number $B = 0$ that contain an equal number of quarks and anti-quarks), as well
as baryon resonances that decay into nucleons (protons or neutrons) and pions. 
Most important, the QCD spectrum does not contain states of isolated quarks or 
gluons, which are instead permanently confined inside hadrons. 

\subsection{Lattice QCD}

The standard formulation of lattice QCD is due to Wilson. He represented the 
gluon field by parallel transporter $N \times N$ unitary matrices $U_{xy}$ of 
determinant 1, that take values in the non-Abelian color gauge group $SU(N)$, 
and are associated with the link connecting nearest neighbor lattice sites $x$ 
and $y$. While Wilson originally constructed the theory in the Lagrangian 
formulation, it was soon expressed by Kogut and Susskind in the Hamiltonian 
formulation \cite{Kog75}. In close analogy to lattice QED, again using staggered
fermions, the lattice QCD Hamiltonian takes the form
\begin{eqnarray}
\label{HamiltonianQCD}
H_{\text{QCD}}&=& 
- t \sum_{\langle x y \rangle} s_{xy} \left(\psi_x^\dagger U_{xy} \psi_y + 
\psi_y^\dagger U_{xy}^\dagger \psi_x \right) + 
m \sum_x s_x \psi_x^\dagger \psi_x \nonumber \\
&+&\frac{g^2}{2} \sum_{\langle x y \rangle} (L_{xy}^2 + R_{xy}^2) - \frac{1}{4 g^2}
\sum_\Box \mbox{Tr} \left(U_\Box + U_\Box^\dagger\right).
\end{eqnarray}
Here we are using one ``flavor'' of staggered fermions with mass $m$. Due
to fermion doubling, in the continuum limit this will give rise to multiple 
fermion species. We have suppressed the color indices, which in a hopping term 
would appear as 
$\psi_x^\dagger U_{xy} \psi_y = {\psi^i_x}^\dagger U^{ij}_{xy} \psi^j_y$. As
in the Abelian case, the plaquette product 
$U_\Box =  U_{wx} U_{xy} U_{zy}^\dagger U_{wz}^\dagger$
represents the color magnetic field. The color electric field is described by 
the flux operators $L_{xy}$ and $R_{xy}$, associated with the left and right end 
of the link $xy$. These non-Abelian analogs of $E_{xy}$ are 
operators that take appropriate derivatives with respect to the matrix elements 
of $U_{xy}$. Suppressing the link index $xy$, the various operators obey the 
commutation relations
\begin{equation}
\label{linkalgebra}
[L^a,L^b] = 2 i f_{abc} L^c, \quad [R^a,R^b] = 2 i f_{abc} R^c, \quad 
[L^a,R^b] = 0, \quad [L^a,U] = - \lambda^a U, \quad [R^a,U] = U \lambda^a.
\end{equation}
Operators associated with different links commute with each other. The Hermitean
generators of $SU(N)$ obey the commutation relation
$[\lambda^a,\lambda^b] = 2 i f_{abc} \lambda^c$, where $f_{abc}$ are the 
structure constants of the $SU(N)$ algebra and
$\mbox{Tr} \lambda^a \lambda^b = 2 \delta^{ab}$. By construction, the 
Hamiltonian of eq.(\ref{HamiltonianQCD}) is gauge invariant, i.e.\ it commutes 
with the infinitesimal generators of $SU(N)$ gauge transformations
\begin{equation}
\label{generatorQCD}
G_x^a = \psi_x^{i \dagger} \lambda^a_{ij} \psi_x^j + 
\sum_k \left(L^a_{x,x+\hat k} + R^a_{x-\hat k,x}\right), \quad
[G^a_x,G^b_y] = 2 i \delta_{xy} f_{abc} G^c_x.
\end{equation}
Again, physical states $|\Psi\rangle$ are gauge invariant and must obey the 
Gauss law $G_x^a |\Psi\rangle = 0$. A general $SU(N)$ gauge transformation, 
$\Omega_x = \exp(i \alpha_x^a \lambda^a)$, is represented by the unitary 
transformation $V = \prod_x \exp(i \alpha_x^a G_x^a)$,
which acts as
\begin{equation}
\label{gaugeQCD}
\psi_x' =  V^\dagger \psi_x V = \Omega_x \psi_x, \quad
{\psi^\dagger_x}' =  V^\dagger \psi^\dagger_x V = 
\psi^\dagger_x \Omega_x^\dagger, \quad
U'_{xy} = V^\dagger U_{xy} V = \Omega_x U_{xy} \Omega_y^\dagger.
\end{equation}

In Wilson's lattice gauge theory, the commutation relations of 
eq.(\ref{linkalgebra}) are realized in an infinite-dimensional Hilbert space 
per link. In fact, every link is analogous to a quantum mechanical ``particle''
moving in the group space $SU(N)$, with $L_{xy}^2 + R_{xy}^2$ representing the
corresponding Laplacian. 

In an $SU(2)$ gauge theory the various operators can be 
represented by harmonic oscillators \cite{Mat05} (also known as prepotentials) 
using bosonic creation and annihilation operators $a^{i \dagger}_{x,\pm k}$ and 
$a^i_{x,\pm k}$, which carry a color index $i \in \{1,2\}$. The bosonic operators 
are associated with the left and right ends of a link and are labeled by 
a lattice point $x$ and a link direction $\pm k$, and one can write
\begin{eqnarray}
L^a_{xy}&=&a^{i \dagger}_{x,+} \sigma^a_{ij} a^j_{x,+}, \
R^a_{xy} = a^{i \dagger}_{y,-} \sigma^a_{ij} a^j_{y,-}, \nonumber \\
U_{xy}&=&\frac{1}{{\cal N}_{xy}} \left(\begin{array}{cc} 
a^{2 \dagger}_{x,+} & a^1_{x,+} \\ - a^{1 \dagger}_{x,+} & a^2_{x,+} 
\end{array}\right)
\left(\begin{array}{cc} a^{1 \dagger}_{y,-} & a^{2\dagger}_{y,-} \\ 
a^2_{y,-} & - a^1_{y,-} \end{array}\right) \nonumber \\
&=&\frac{1}{{\cal N}_{xy}} \left(\begin{array}{cc} 
a^{2 \dagger}_{x,+} a^{1 \dagger}_{y,-} + a^1_{x,+} a^2_{y,-} &
a^{2 \dagger}_{x,+} a^{2 \dagger}_{y,-} - a^1_{x,+} a^1_{y,-} \\
- a^{1 \dagger}_{x,+} a^{1 \dagger}_{y,-} + a^2_{x,+} a^2_{y,-} &
- a^{1 \dagger}_{x,+} a^{2 \dagger}_{y,-} - a^2_{x,+} a^1_{y,-} 
\end{array}\right).
\end{eqnarray}
Here ${\cal N}_{xy} = a^{i \dagger}_{x,+} a^i_{x,+} = a^{i \dagger}_{y,-} a^i_{y,-}$ 
counts the number of bosons, which is the same at both ends of the link. The 
link operator $U_{xy}$ changes the number of bosons by two, by either
creating or annihilating a boson on each end of a link. Since the link Hilbert
space is infinite-dimensional, the total number of bosons can be arbitrarily
large. In $SU(3)$ gauge theory the construction is much more involved 
\cite{Ani10}. Instead of two, it involves four species of colored bosons per 
link, which span a Hilbert space that is larger than the one of the gauge 
theory. In order to correct for this, the link operators are no longer 
constructed as boson bilinears, but as polynomials of a higher degree. Even 
then, the commutation relations of eq.(\ref{linkalgebra}) are satisfied only in 
the gauge theory subspace of the bosonic Hilbert space, and it is not obvious 
how to restrict oneself to that subspace. While there are constructions for
quantum simulators using $SU(2)$ prepotentials \cite{Zoh13a,Zoh13b}, it is 
difficult to imagine that the $SU(3)$ prepotentials of \cite{Ani10} can be 
implemented in ultracold matter.

Up to now, we have defined the theory on a lattice with non-zero lattice spacing
$a$, whose inverse $\frac{1}{a}$ serves as an ultraviolet momentum cut-off.
Ultimately, we want to take the continuum limit $a \rightarrow 0$. This is 
done by properly adjusting the bare coupling constant $g$. We may fix the 
overall energy scale by putting $t = 1$. When we set $m = 0$, we are in the 
chiral limit of massless quarks, which will lead to a massless Goldstone pion. 
The bare gauge coupling $g$ is then adjusted in order to take the continuum 
limit. This can be done by considering any dimensionful physical quantity, for 
example, the nucleon mass. The nucleon mass $M_n = E_1 - E_0$ is the energy 
difference between the ground states of $H_{\text{QCD}}$ in the baryon number 1 
and 0 sectors. Let us 
consider the nucleon mass in lattice units, i.e.\ $M_n a$, as a function of $g$.
Due to the property of asymptotic freedom, in the $g \rightarrow 0$ limit the 
nucleon mass behaves as $M_n a \sim \exp(- \beta_0/g^2)$, where $\beta_0 > 0$ is
the leading coefficient of the QCD $\beta$-function. Keeping the physical 
quantity $M_n$ fixed and sending $g \rightarrow 0$, one approaches the 
continuum limit $a \rightarrow 0$. We have thus traded the dimensionless bare 
coupling constant $g$ for a dimensionful physical scale --- in this case $M_n$. 
In this process of dimensional transmutation, the scale invariance of the QCD 
Lagrangian in the massless chiral limit is explicitly broken by the ultraviolet 
regulator $\frac{1}{a}$. It should be pointed out that massless QCD does not 
predict the value of any dimensionful scale like the nucleon mass. After all, 
the nucleon mass, e.g.\ in units of kilograms, relies on a man-made convention, 
and essentially reduces to the question how many protons and neutrons were 
deposited near Paris, when the kilogram was defined a long time ago. However, 
once an overall energy scale, e.g.\ $M_n$, has been picked, QCD predicts the 
values of countless dimensionless ratios, including the masses of nuclei 
$M_B = E_B - E_0$ in units of $M_n$. Here $E_B$ is the energy of the ground 
state in the sector with baryon number $B$, for example, the deuteron bound 
state of a proton and a neutron for $B = 2$.

QCD gives rise to rich dynamics both at non-zero temperature and at
non-zero baryon density. While chiral symmetry is spontaneously broken at low
temperatures, it is restored at the high temperatures that existed in the early
universe, and that are recreated in heavy-ion collisions. The crossover that
separates the low- from the high-temperature phase has been accurately
investigated using Monte Carlo simulations of lattice QCD \cite{Aok06,Baz09}.
The QCD physics of 
non-zero baryon number, which governs individual nuclei as well as the matter 
deep inside neutron stars, is much less well understood from first principles.
This is because Monte Carlo simulations fail at non-zero baryon density, due to
a very severe sign problem. The same is true for studies of the real-time 
evolution, e.g.\ of a heavy-ion collision. These dynamics as well as the phases
of matter at high baryon densities, which may include color superconductors 
\cite{Raj00,Alf08} (with or without color-flavor locking) provide a strong 
motivation to construct quantum simulators for non-Abelian gauge theories.

\subsection{Non-Abelian Quantum Link Models}

In contrast to Wilson's lattice gauge theory, in quantum link models the real 
and imaginary parts of $U_{xy}^{ij}$ are no longer real numbers, but 
non-commuting Hermitean operators. The commutation relations of 
eq.(\ref{linkalgebra}) are then realized by using the generators of an 
$SU(2N)$ algebra. The real and imaginary parts of the $N^2$ matrix
elements $U_{xy}^{ij}$ are represented by $2N^2$ Hermitean generators of the 
embedding algebra $SU(2N)$. The $2(N^2 - 1)$ generators $L^a_{xy}$ and $R^a_{xy}$ 
of $SU(N)$ gauge transformations at the left and at the right end of the link 
also belong to the $SU(2N)$ algebra. An additional Abelian $U(1)$ gauge 
transformation (which does not distinguish between left and right) is
related to the generator $E_{xy}$. Altogether there are 
$2 N^2 + 2 (N^2 - 1) + 1 = 4 N^2 - 1$ generators, which form the embedding
algebra $SU(2N)$. Similar constructions exist for $SO(N)$ and $Sp(N)$ gauge
theories, which are naturally represented with $SO(2N)$ and $Sp(2N)$ embedding
algebras \cite{Bro04}. Unlike in Wilson's lattice gauge theory, where the 
different elements of a link matrix commute with each other as well as with 
their complex conjugates, i.e.\ $[U^{ij},U^{kl\dagger}] = 0$, the elements of a 
quantum link matrix, $U^{ij}$, are non-commuting operators that obey the 
commutation relations
\begin{equation}
\label{qlinks}
[U^{ij},U^{kl\dagger}] = 
2 (\delta_{ik} \sigma^{a *}_{jl} R^a - \delta_{jl} \sigma^a_{ik} L^a +
2 \delta_{ik} \delta_{jl} E), \quad
[U^{ij},U^{kl}] = [U^{ij\dagger},U^{kl\dagger}] = 0.
\end{equation}
Again, commutators between operators assigned to different links are zero. 
All other relations of Wilson's QCD, including the Hamiltonian of 
eq.(\ref{HamiltonianQCD}) and the gauge transformations of 
eqs.(\ref{generatorQCD}) and (\ref{gaugeQCD}), remain valid in quantum link 
modes. The advantage of this alternative formulation of non-Abelian gauge 
theories is again that the Hilbert space is now finite-dimensional. Just as we 
could pick any value of the quantum spin $S$ in the Abelian $U(1)$ quantum link
model (with the embedding algebra $SU(2)$), in the non-Abelian $SU(N)$ quantum
link model we can pick any representation of the embedding algebra $SU(2N)$. In 
the appropriate limit of a large representation, one recovers the Wilson theory 
\cite{Sch01}. However, quantum link models are more than just a gauge invariant 
truncation of the Wilson theory. Indeed, they provide a non-trivial 
generalization of the concept of lattice gauge theories, with ample room for 
interesting new physical phenomena at strong coupling. As we have seen 
explicitly for the $(2+1)$-d $U(1)$ quantum link model, this also provides 
interesting connections to condensed matter physics. 

It turns out that, in addition to the non-Abelian $SU(N)$ gauge invariance, 
the Hamiltonian of eq.(\ref{HamiltonianQCD}) also has an Abelian $U(1)$ 
gauge symmetry which is generated by
\begin{equation}
G_x = \psi_x^{i \dagger} \psi_x^i +
\sum_k \left(E_{x,x+\hat k} - E_{x-\hat k,x}\right), \ [G_x,G_x^a] = 0.
\end{equation}
If one wants to construct an $SU(N)$ rather than a $U(N)$ gauge theory, one
should explicitly break the additional $U(1)$ symmetry. This can be
done by adding the term $\gamma \sum_{\langle x y \rangle} 
\left(\mathrm{det} U_{xy} + \mathrm{det} U_{xy}^\dagger\right)$ to the Hamiltonian.
Since all matrix elements $U^{ij}_{xy}$ commute with each other (although 
$U^{ij}_{xy}$ and $U^{kl\dagger}_{xy}$ do not commute) the definition of 
$\mbox{det} U_{xy}$ does not suffer from operator ordering ambiguities. 

The commutation relations of eqs.(\ref{linkalgebra}) and (\ref{qlinks}) can be 
realized using fermionic rishons, representing constituents of gauge bosons. 
The rishon operators $c^i_{x,\pm k}$, $c^{i \dagger}_{x,\pm k}$ with color index 
$i \in \{1,2,\dots,N\}$ lead to the representation
\begin{equation}
L^a_{xy} = c^{i \dagger}_{x,+} \lambda^a_{ij} c^j_{x,+}, \
R^a_{xy} = c^{i \dagger}_{y,-} \lambda^a_{ij} c^j_{y,-}, \
E_{xy} = \frac{1}{2}(c^{i \dagger}_{y,-} c^i_{y,-} - c^{i \dagger}_{x,+} c^i_{x,+}), \
U^{ij}_{xy} = c^i_{x,+} c^{j \dagger}_{y,-}.
\end{equation}
Rishon operators are associated with the left and right ends of a link and 
are labeled with a lattice point $x$ and a link direction $\pm k$. They obey
canonical anti-commutation relations
\begin{equation}
\{c^i_{x,\pm k},c^{j \dagger}_{y,\pm l}\} = \delta_{xy} \delta_{\pm k,\pm l} \delta_{ij}, 
\ \{c^i_{x,\pm k},c^j_{y,\pm l}\} = \{c^{i \dagger}_{x,\pm k},c^{j \dagger}_{y,\pm l}\} = 0.
\end{equation}
Quark and rishon operators also anti-commute with each other. All operators 
introduced so far (including the Hamiltonian) commute with the rishon number 
operator ${\cal N}_{xy} = c^{i \dagger}_{x,+} c^i_{x,+} + c^{i \dagger}_{y,-} c^i_{y,-}$
on each individual link. Hence, we can limit ourselves to sectors of fixed 
rishon number for each link. This is equivalent to working in a given 
irreducible representation of $SU(2N)$. We can write the determinant operator 
that we used to break the $U(N)$ gauge symmetry down to $SU(N)$ as
\begin{equation}
\mbox{det} U_{xy} = \frac{1}{N!} \epsilon_{i_1 i_2 ... i_N} 
U_{xy}^{i_1 i'_1} U_{xy}^{i_2 i'_2} \dots U_{xy}^{i_N i'_N} \epsilon_{i'_1 i'_2 \dots i'_N} =
N! \ c^1_{x,+} c^{1 \dagger}_{y,-} c^2_{x,+} c^{2 \dagger}_{y,-} \dots 
c^N_{x,+} c^{N \dagger}_{y,-}.
\end{equation}
Only when this operator acts on a state with exactly ${\cal N} = N$ rishons
(all of a different color), it gives a non-zero contribution. In all other 
cases the determinant vanishes. This means that we can reduce the symmetry from
$U(N)$ to $SU(N)$ via the determinant when we work with ${\cal N} = N$ 
fermionic rishons on each link. The number of fermion states per
link then is $(2N)!/(N!)^2$. For an $SU(2)$ gauge theory the Hilbert space of
the corresponding quantum link model then has 6 states per link, while for 
$SU(3)$ it has 20 states per link. In the standard Wilson formulation of 
lattice gauge theory, on the other hand, the single link Hilbert space is 
already infinite-dimensional. This limit can be recovered when one introduces a
large number of rishon ``flavors'' \cite{Sch01}. The rishon dynamics are 
illustrated in Figure 5a.
\begin{figure}[tbp]
\includegraphics[width=0.46\textwidth]{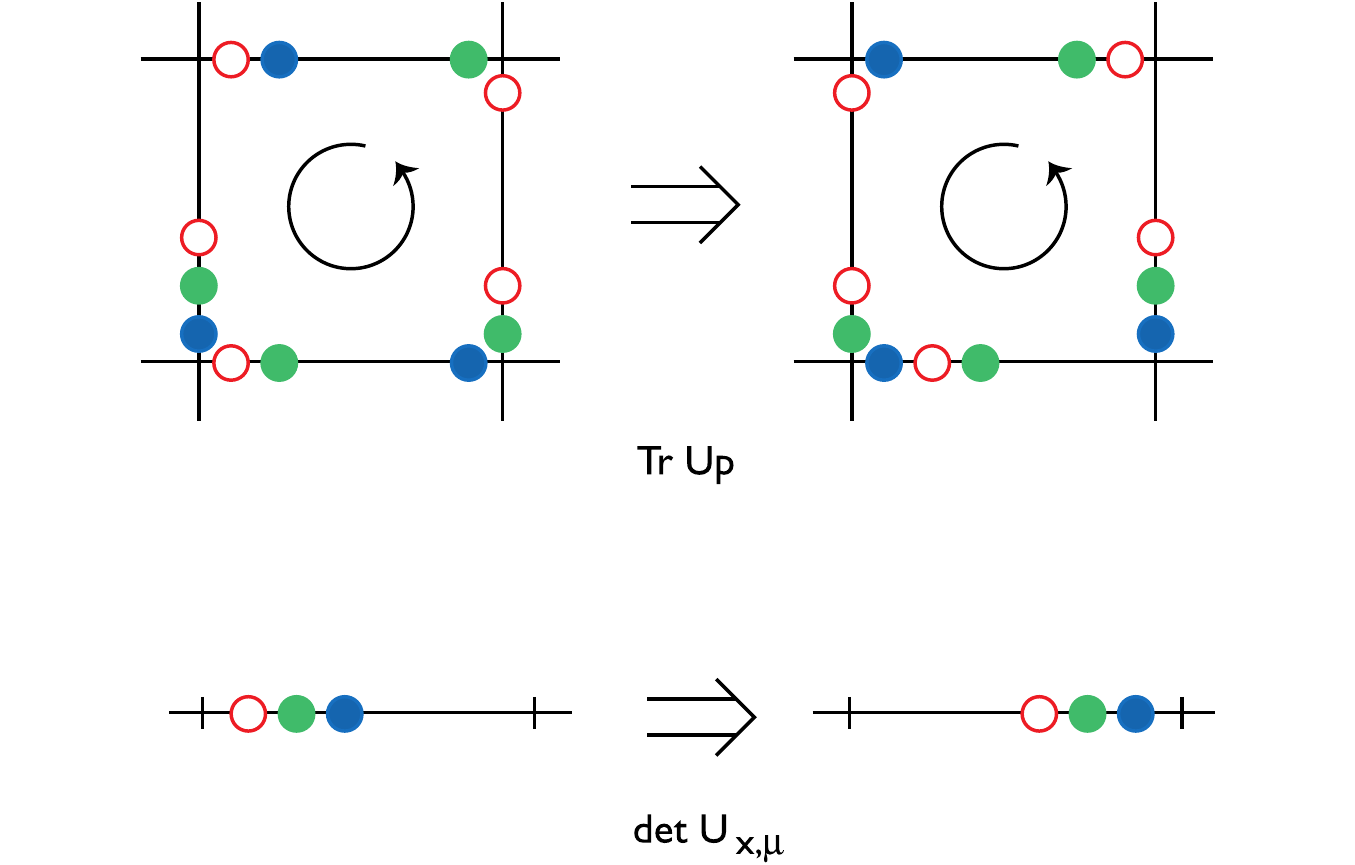}
\includegraphics[width=0.54\textwidth]{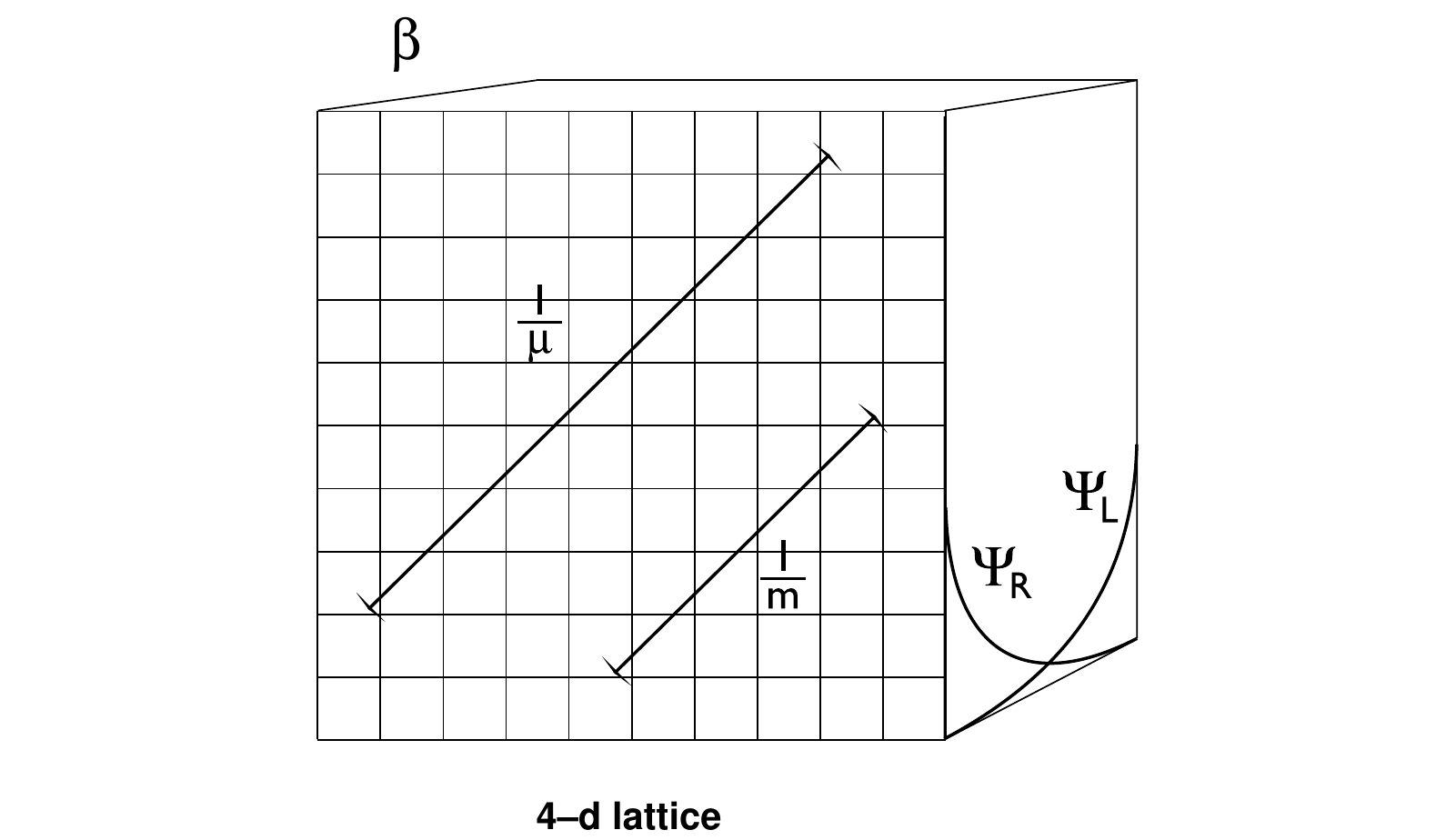}
\caption{\textit{ Left: a) In an $SU(3)$ quantum link model, three 
fermionic rishons of different colors reside on each link. The Hamiltonian 
moves the rishons around plaquettes or along links, like the beads of an abacus.
Right: b) In the quantum link formulation of QCD, the size $\beta$ of an extra 
dimension is small compared to the gluonic and fermionic correlation lengths 
$1/m$ and  $1/\mu$, and the theory undergoes dimensional reduction. Effective 
continuous gluon fields emerge from the discrete quantum link variables, and 
quark modes $\Psi_L$ and $\Psi_R$ appear as domain wall fermions.}}
\end{figure}

In the D-theory formulation of quantum field theory, the continuum limit of 
quantum link models is approached by dimensional reduction from a higher 
dimension \cite{Cha97}. Since  
non-Abelian gauge theories in $(4+1)$-d have massless Coulomb phases, one 
naturally reaches the exponentially small masses of asymptotically free field
theories by dimensional reduction. Effective continuous 4-d gluon fields then 
arise as collective excitations of discrete quantum link variables, just as 
effective magnon fields arise as collective excitations of quantum spins. 
Chiral quarks are incorporated naturally as domain wall fermions, which arise 
as massless 4-d edge-modes on a $(4+1)$-d slab, analogous to the edge-states of 
a quantum Hall sample (c.f.\ Figure 5b). Universality then implies that
one recovers QCD in the continuum limit.

Quantum simulators will first be realized in the strong coupling regime, in 
which there is no need for large link Hilbert spaces or complicated plaquette 
interactions. Even if they do not yet represent Nature's QCD, strongly coupled
lattice gauge models already display rich dynamics including confinement and 
deconfinement, chiral symmetry breaking and restoration , and perhaps even 
color superconductivity. By quantum simulating such models, one may learn a lot 
about fundamental dynamical mechanisms in non-Abelian gauge theories, which 
will be very useful both in particle and in condensed matter physics. 
 
\section{Quantum Simulators for non-Abelian Lattice Gauge Theories}

In this section, we discuss both digital and analog quantum simulators for
non-Abelian gauge theories, again with and without matter fields.

\subsection{Digital Quantum Simulator for an $SU(2)$ Gauge Theory}

Because $SU(2) \simeq SO(3)$ is also equivalent to $Sp(1)$, it can be realized 
both in an $SU(4) \simeq SO(6)$ and in an $Sp(2) \simeq SO(5)$ embedding
algebra. The quantum link model that was originally constructed by Horn is based
on an $SO(5)$ 
embedding \cite{Hor81}. Orland and Rohrlich used the 4-state $SO(5)$ spinor 
representation in their study of the $(2+1)$-d gauge magnet \cite{Orl90}. A 
quantum simulator for this system, using Rydberg atoms in an optical lattice, 
has been proposed in \cite{Tag13}. It utilizes two ensemble atoms per link, in 
a holographically designed optical lattice \cite{Bak09,Wei11}, which encode the 
4 link states, as well as control atoms at the lattice sites and at the 
plaquette centers. These atoms control an eight-qubit interaction realizing the 
plaquette product $U_\Box$. Protocols for ground state preparation in the 
presence of external static charges, as well as methods for detecting the 
confining strings connecting the charges by direct imaging have also been 
proposed.

\subsection{Analog Quantum Simulators for $U(N)$ and $SU(N)$ Gauge Theories}

Analog quantum simulators for $SU(2)$ gauge theories coupled to fermionic 
matter, based on the bosonic prepotential representation of Wilson's $SU(2)$ 
gauge theory \cite{Mat05} have been proposed in \cite{Zoh13a,Zoh13b} using a 
Fermi-Bose mixture of atoms in an optical superlattice. Four bosonic and one
auxiliary fermionic species reside on each link, representing the ``gluon''
gauge field. In order to model the infinite-dimensional Hilbert space of the 
Wilson theory, two bosonic species form a Bose-Einstein condensate. The 
``quarks'' are embodied by another species of fermionic atoms. Using 
higher-order processes, plaquette interactions have also been constructed
in this way.

\begin{figure}[t]
\begin{center}
\includegraphics[width=0.45\textwidth]{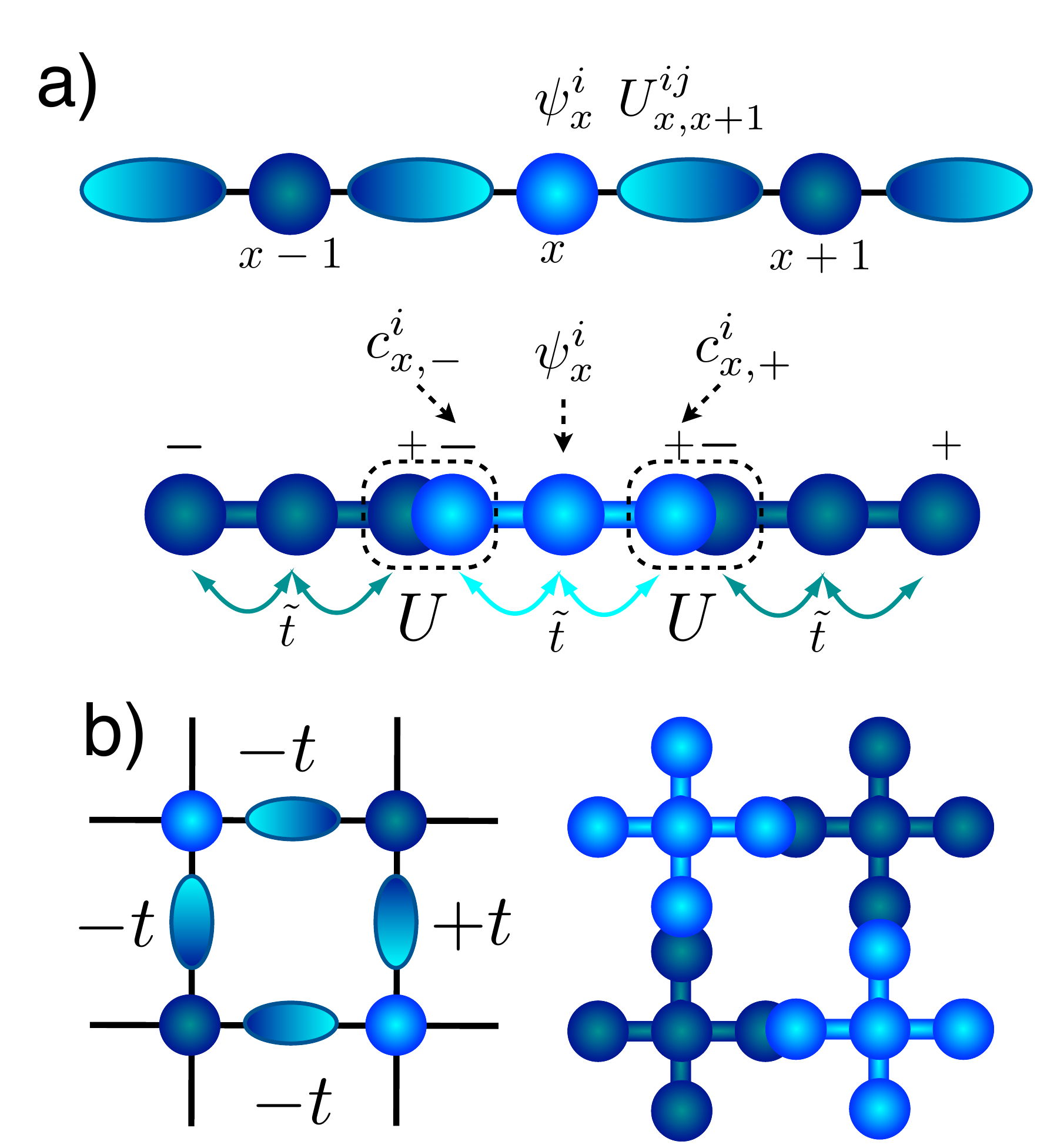} \hspace{0.3cm}
\includegraphics[width=0.45\textwidth]{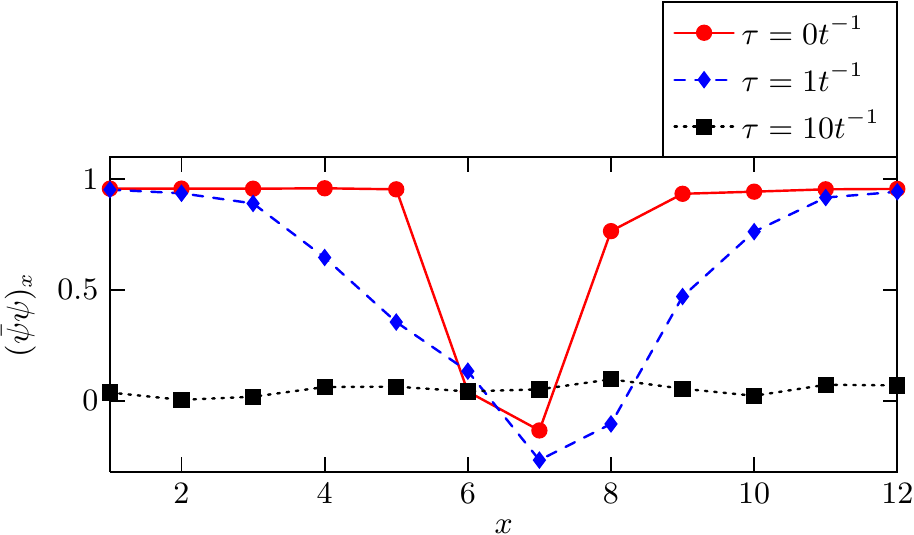}
\caption{\it Left: Quantum simulator for an $SU(N)$ gauge theory with ultracold 
alkaline-earth atoms representing ``quarks'' or rishon constituents of 
``gluons'' in a 1-d (a) or 2-d (b) optical superlattice. Right: c) Real-time 
$\tau$ evolution of the chiral order parameter profile $(\overline \psi \psi)_x$
in a $U(2)$ gauge theory, mimicking the expansion of a hot quark-gluon plasma
\cite{Ban13}.}
\end{center}
\end{figure}
An alternative proposal is based on the rishon formulation of $U(N)$ and
$SU(N)$ quantum link models \cite{Ban13}. This construction uses fermionic
alkaline-earth atoms \cite{Caz09,Gor10,Tai12}, such as $^{87}Sr$ or $^{173}Yb$, 
hopping in an optical superlattice \cite{Dal08}. The interactions between the
atoms can be engineered, e.g., using Feshbach resonances \cite{Ciu05,Chi10}.
The color degree of freedom is encoded in 
the Zeeman states of the nuclear spin $I$, which enjoy an $SU(2I+1)$ symmetry.
Each link has two rishon-sites, one associated with each end of the link. The 
number of rishons,
${\cal N}_{xy} = c_{x,+k}^{i\dagger} c_{x,+k}^i + c_{y,-k}^{i\dagger} c_{y,-k}^i$, on each
link is fixed to $n$, due to a Hubbard-type energy penalty
\begin{equation}
\widetilde{H} = \tilde t \sum_{x,k} \left(s_{xy} c^{i \dagger}_{x,+k} \psi^i_x + 
c^{i \dagger}_{x,-k} \psi^i_x + \mathrm{h.c.}\right) + 
m \sum_x s_x \psi_x^{i\dagger} \psi^i_x + 
U \sum_{\langle xy \rangle}(\mathcal{N}_{xy} - n)^2.
\end{equation}
When an atom resides on a rishon-site $x,\pm k$ it embodies a fermionic 
constituent of a ``gluon''. When the same atom hops to an ordinary lattice site
$x$, it embodies a ``quark''. $SU(N)$ gauge invariance is thus naturally 
protected by the dynamics, while the additional $U(1)$ gauge symmetry is 
explicitly violated at the cut-off scale, but emerges at low energy.
Small explicit violations of the gauge symmetry are tolerable, because gauge
invariance is robust at low energies \cite{Foe80,Kas13}. $U(N)$ gauge
theories do not contain baryons, because $U(1)$ is gauged, and baryons are
confined at strong coupling. $U(2)$ lattice gauge theories, which can be 
realized with a single rishon per link, are still interesting, e.g., because 
they have a spontaneously broken chiral symmetry, whose dynamics can be
investigated in real time. Figure 6c shows the evolution of the profile
of the chiral condensate $(\overline\psi \psi)_x$ (i.e.\ the order parameter for
spontaneous chiral symmetry breaking), derived by exact diagonalization of a 
moderate-size 1-d system. The initial configuration contains a 
small region in which the order parameter deviates from its vacuum value, thus
mimicking a hot quark-gluon plasma. As time evolves, this region expands. When 
${\cal N} = N$ one can use the $\mbox{det}U$-term to reduce the gauge symmetry 
from $U(N)$ to $SU(N)$. For 
$SU(2)$ one then obtains bosonic ``baryons'', whose quantum simulation requires
simultaneous two-rishon tunneling, which can be implemented as a Raman process.
In such a model, one can study chiral symmetry restoration at non-zero baryon
density. Nature's $SU(3)$ gauge symmetry yields fermionic baryons (e.g., protons
and neutrons) which would require the engineering of a three-rishon tunneling 
process (c.f.\ Figure 5a, bottom), whose realization is more challenging.

\section{Conclusions and Outlook}

As we have seen, there are many exciting ideas how to quantum simulate dynamical
lattice gauge theories, which are of interest both in particle and in condensed
matter physics. When quantum link models or gauge magnets with plaquette
interactions are used to model condensed matter systems, such as spin liquids, 
one can use digital simulations using Rydberg atoms in an optical lattice. It
would also be very interesting to investigate whether $(2+1)$-d Chern-Simons
gauge theories can be regularized with quantum link models. If so, one might be
able to realize non-Abelian anyons with ultracold matter in optical lattices.
This might enable topological quantum computation \cite{Nay08}, which is robust 
against
decoherence, with exquisite experimental control. Aiming at particle physics 
applications, one may pursue different strategies. When one uses the Wilson 
formulation with bosonic prepotentials, the continuous nature of Wilson's 
parallel transporters, which results in an infinite-dimensional link Hilbert 
space, demands a large number of bosons per link, possibly forming a 
Bose-Einstein condensate. The continuum limit is then taken by tuning the 
plaquette coupling, which is non-trivial to engineer, to a large value (which 
corresponds to sending $g \rightarrow 0$). Plaquette couplings 
can also be induced with additional fermion flavors \cite{Has92}, or with
scalar fields \cite{Wet12}.

Quantum link models offer an alternative to this strategy. Thanks to the 
discrete nature of the quantum link variables, these models can be embodied by a
few particles per link, thus benefiting from their low-dimensional link Hilbert 
space. Generically, quantum link models are lattice gauge theories at strong 
coupling, which often do not need plaquette interactions to display interesting 
dynamics. As such, they offer a unique opportunity to explore fundamental 
features of strong gauge dynamics with a minimal set of degrees of freedom. 
This includes investigations of confinement and deconfinement, chiral symmetry 
breaking and restoration at finite baryon density \cite{Cha06,Cha10}, and 
perhaps even color superconductivity. Long before one ultimately takes the 
continuum limit, quantum simulating these intriguing phenomena with ultracold 
atomic gases would be most exciting, and indeed seems within reach in the 
foreseeable future.

Various types of Bose-Bose, Fermi-Fermi and Bose-Fermi mixtures are available 
in the laboratory. Different set-ups may be engineered within state of the art 
technologies \cite{Blo08}, each of them highly tunable and characterized by 
efficient detection techniques such as in situ imaging, leading to precise
snapshots of the particles' spatial distribution. Ultracold gases of 
heteronuclear molecules offer even more tunable set-ups, since one can exploit 
the rich internal structure of such objects to properly design several types of 
interactions \cite{Lah09}. Rydberg atoms in optical lattices or alkaline-earth 
atoms offer further opportunities. While the realization of the proposed 
quantum simulators is challenging, and will require the combination of several 
existing experimental techniques, the ball is now in the court of the 
experimental colleagues, to utilize their rich arsenal of powerful experimental 
techniques, and to go ahead and quantum simulate dynamical gauge fields. 
Theorists should support this process, by further simplifying the existing 
constructions, wherever this seems possible. Experimentally, one probably wants 
to start with a simple case. A well-defined first goal could be to realize the 
$(1+1)$-d $U(1)$ quantum link model with fermionic matter, in order to study 
the real-time evolution of string breaking. Moving on to higher dimensions and 
to non-Abelian gauge theories are natural next steps.

An ultimate long-term goal will be to quantitatively address the QCD phase 
diagram at non-zero baryon density and the real-time evolution of heavy-ion
collisions. This will require a correct implementation of chiral symmetry. In
the quantum link model based D-theory formulation of QCD, this is 
naturally provided by domain wall fermions, which requires multi-component
fermion fields for which further tools from atomic physics would have to be 
developed. In D-theory, the continuum limit of an asymptotically free gauge 
theory, such as QCD, is taken by the dimensional reduction of the discrete 
quantum link variables from an additional spatial dimension, with an extent of 
just a few lattice spacings. While it will be challenging to endow ultracold 
matter with something equivalent to an extra dimension, this seems not 
impossible. In particular, one can imagine encoding the discrete coordinate of 
the extra dimension in an internal atomic state or realize it  with distinct 
atomic species. Recovering Lorentz invariance will happen almost automatically 
in the D-theory set-up, except that the ``velocity of light'' of the quarks 
must be tuned to match the one of the gluons. While quantum simulating QCD near 
the continuum limit may remain a very long-term goal, there is a lot of 
exciting physics to be discovered along the way.

\section*{Acknowledgments}

Quantum link models were developed as an alternative non-perturbative 
regularization of Yang-Mills theory together with S.\ Chandrasekharan and 
extended to full QCD with R.\ Brower. I thank both of them for their 
collaboration and long friendship. The idea that quantum links are ideally 
suited for quantum simulation of dynamical gauge theories occurred to me at a 
KITP workshop in Santa Barbara in 2010. I thank J.\ I.\ Cirac and M.\ Lewenstein
for encouraging me at that time to pursue this idea. I'm grateful to H.\ Briegel
and M.\ A.\ Martin-Delgado for inviting me to the ``Quantum Information meets
Statistical Mechanics'' summer school at El Escorial in 2011, which brought me 
in touch with P.\ Zoller and his group, including M.\ Dalmonte, M.\ M\"uller,
and E.\ Rico Ortega. Several results reviewed here resulted from 
a very pleasant and fruitful collaboration with them, which I very much hope 
will continue for a long time in the future. I also like to thank D.\ Banerjee, 
M.\ B\"ogli, F.-J.\ Jiang, P.\ Stebler, and P.\ Widmer for close collaboration 
on many of the topics discussed here. Last but not least, I like to thank P.\
Zoller for inviting me to write this review.

\end{document}